\documentclass[reprint,a4paper]{article}
\usepackage{eurosym}
\usepackage[affil-it]{authblk}
\usepackage{setspace}
\usepackage{color}
\usepackage[dvipsnames]{xcolor}
\usepackage{amsmath}
\usepackage{amsfonts}
\usepackage{subcaption}
\usepackage[title]{appendix}
\usepackage{comment}
\usepackage{booktabs}
\usepackage{verbatim}
\usepackage{amssymb}
\usepackage{graphicx}
\usepackage{hyperref}
\hypersetup{
    colorlinks=true,
    linkcolor=blue,
    filecolor=magenta,
    urlcolor=[rgb]{0.00,0.00,0.50},
    citecolor=red,
    }
\usepackage[nameinlink]{cleveref}
\usepackage{cite}
\usepackage{bbold}
\usepackage{dsfont}
\usepackage{url}
\usepackage{caption}
\usepackage{subcaption}
\usepackage[top=2cm, bottom=2cm, left=3cm, right=3cm]{geometry}
\usepackage{upgreek}

\newcommand{\ii}{\ensuremath{\mathrm{i}}}
\definecolor{darkgreen}{rgb}{0,0.35,0}

\newcommand{\CECs}{Centro de Estudios Cient\'{\i}ficos (CECs), Casilla 1469, Valdivia, Chile}
\newcommand{\USS}{Universidad San Sebasti\'{a}n, sede Valdivia, General Lagos 1163, Valdivia 5110693, Chile}
\newcommand{\UACh}{Instituto de Ciencias F\'isicas y Matem\'aticas, Universidad Austral de Chile, Casilla 567, 5090000 Valdivia, Chile}
\newcommand{\UCharles}{IPNP - Faculty of Mathematics and Physics, Charles University, V Hole\v{s}ovi\v{c}k\'ach 2, 18000 Prague 8, Czech Republic}

\begin{document}

\title{A Bogomol'nyi-Prasad-Sommerfield bound with a first-order system in the $2D$ Gross-Pitaevskii equation}
\author[1,2]{Fabrizio Canfora\thanks{fabrizio.canfora@uss.cl}}
\author[3,4]{Pablo Pais\thanks{pais@ipnp.troja.mff.cuni.cz}}

\affil[1]{\CECs} \affil[2]{\USS} \affil[3]{\UACh} \affil[4]{\UCharles}

\date{}
\maketitle

\begin{abstract}
A novel Bogomol'nyi-Prasad-Sommerfield (BPS) bound for the Gross-Pitaevskii equations in two spatial dimensions is presented. The energy can be bounded from below in terms of the combination of two boundary terms, one related to the vorticity (but ``dressed'' by the condensate profile) and the second to the ``skewness'' of the configurations. The bound is saturated by configurations that satisfy a system of two first-order partial differential equations. When such a BPS system is satisfied, the Gross-Pitaevskii equations are also satisfied. The analytic solutions of this BPS system in the present manuscript represent configurations with fractional vorticity living in an annulus. Using these techniques, we present the first analytic examples of this kind. The hydrodynamical interpretation of the BPS system is discussed, and the implications of these results are outlined.
\end{abstract}

%%%%%%%%%%%%%%%%%%%%%%%%%%%%%%%%%%%%%%%%%%%%%%%%%%%%%%%%%%%%%%%%%%%%%%%%%%%%%%%%%%%%%%%%%%%%%%%%%%%%%%%%%%%%%%%%%%%%%%%%%%%%%%%%%%%%%%%%%%%%%%%%%%%%%%%%%

%%%%%%%%%%%%%%%%%%%%%%%%%%%%%%%%%%%%%%%%%%%%%%%%%%%%%%%%%%%%%%%%%%%%%%%%%%%%%%%%%%%%%%%%%%%%%%%%%%%%%%%%%%%%%%%%%%%%%%%%%%%%%%%%%%%%%%%%%%%%%%%%%%%%%%%%%

%%%%%%%%%%%%%%%%%%%%%%%%%%%%%%%%%%%%%%%%%%%%%%%%%%%%%%%%%%%%%%%%%%%%%%%%%%%%%%%%%%%%%%%%%%%%%%%%%%%%%%%%%%%%%%%%%%%%%%%%%%%%%%%%%%%%%%%%%%%%%%%%%%%%%%%%%

%%%%%%%%%%%%%%%%%%%%%%%%%%%%%%%%%%%%%%%%%%%%%%%%%%%%%%%%%%%%%%%%%%%%%%%%%%%%%%%%%%%%%%%%%%%%%%%%%%%%%%%%%%%%%%%%%%%%%%%%%%%%%%%%%%%%%%%%%%%%%%%%%%%%%%%%%

\newpage

%%%%%%%%%%%%%%%%%%%%%%%%%%%%%%%%%%%%%%%%%%%%

\section{Introduction}

\label{sec-1} %%%%%%%%%%%%%%%%%%%%%%%%%%%%%%%%%%%%%%%%%%%%

The Gross-Pitaevskii equation (GPE henceforth) is, without doubt, one of the
most important systems of non-linear partial differential equations (PDE
henceforth) in theoretical physics. The tremendous success of the GPE in describing many non-trivial experimental features of superfluids and supersolids is well recognized (see \cite{R0,R01,R11,R2,R2a,sign1,sign2,sign3m,GPE2,GPE3,GPE4,GPE5,1suso,2suso,3suso,4suso,5suso,6suso,7susoa,7susob,7susoc,8suso,9suso,10suso,11suso,12suso,13suso,14suso}
and references therein). The power of the GPE can be compared with the
effectiveness of Ginzburg-Landau free energy in describing superconductors.
As in superconductors, quantized vortices in superfluids play a fundamental
role both in determining the equilibrium and out-of-equilibrium properties,
the primary tool to study them being, obviously, the GPE (see \cite%
{IsospinR1,Brandt}).

However, unlike what happens in the Ginzburg-Landau theory for
superconductors, unless one is interested in the GPE in one spatial
dimension (which is integrable), there are very few effective analytic tools
to study the relevant non-perturbative configurations of the GPE (such as
quantized vortices). The present manuscript aims to fill this gap.

Of course, an obvious question is: \textit{why should one insist on finding
novel analytic methods and solutions if the GPE can be solved numerically?}
Indeed, the references mentioned here above show in a very clear way that
effective numerical techniques to analyze many statical and dynamical
features of the GPE are already well known. In fact, despite the existence
of many numerical techniques, there are many compelling physical arguments
pushing to search for novel analytic techniques nevertheless. \textit{Firstly%
}, many fundamental concepts have been disclosed and clarified\footnote{%
Much of what we now know about black hole physics in general relativity,
non-perturbative effects in non-Abelian gauge theory, and vortex dynamics
(as well as transitions from type I to type II behaviours in
superconductors) arose from analytic tools.} thanks to the availability of
exact analytic solutions in gauge theory, general relativity, the theory of
superconductors and so on. \textit{Secondly}, there are many open problems
in superfluids in general and in the theory of GPE in particular, where even
numerical methods are not especially effective (such as the transition to
chaos in quantum turbulence: see \cite{GPE5} and references therein). Thus,
the development of analytic tools is not just of academic interest as there
are relevant physical properties that could not be discovered with numerical
tools, as we will see in the following sections.

The most effective technique to analyze topological solitons such as
vortices in superconductors is related to the theory of the
Bogomol'nyi--Prasad--Sommerfield (henceforth BPS) bounds for the (free)
energy of the configurations of interest in terms of the relevant
topological charges (which, in the case of vortices in superconductors, is
the magnetic flux). There are special points in parameter space where it
becomes possible to saturate the BPS bound. The configurations that saturate
these bounds are called BPS solitons. Their relevance can be easily
understood by considering that BPS solitons minimize the (free) energy in
their corresponding topological sectors, a fact that ensures their stability.

A further relevant property is that the saturation of the BPS bound (which
leads to a first-order system of differential equations) implies that the
second-order field equations are also satisfied. This is a significant
property not only because the BPS equations are easier to solve than the
complete set of field equations. Due to their first-order nature, the BPS
system provides a powerful tool to study the low energy dynamics of BPS
configurations through the geodesics of the moduli-space geometry (see \cite%
{modulia,modulib,moduli2,moduli3} and references therein). Moreover, these
BPS points signal the transition between very different behaviors (for
instance, in the Ginzburg-Landau theory of superconductivity, the critical
coupling signals the transition from type I to type II behavior). Last but
not least, the analysis of fermions propagating in the background of BPS
solitons is simplified by different index theorems (see \cite{WeinbergBook}
and references therein). Hence, the appearance of special BPS points in
parameter space is not relevant just due to the fact that the second-order
field equations reduce to a first-order system (one may observe that, for
instance, the BPS equations for the vortices in the Ginzburg-Landau theory
for superconductors at critical coupling are not solvable analytically). All
the above arguments clearly show that BPS points (when they exist) are
extremely important due to their highly effective non-perturbative results,
which allow the analysis of the static and dynamic effects mentioned above.

Until very recently, neither BPS bounds nor BPS equations were available in
the theory of superfluids and GPE in two (or three) spatial dimensions. It
is believed that the GPE in two or more spatial dimensions does not possess
a first-order BPS system with the property that when the BPS system is
satisfied, the GPE is also satisfied. This belief arises from the fact that
the obvious BPS bound (where the vorticity appears on the right-hand side of
the bound) cannot be saturated.

In fact, in the present manuscript, we will construct the first example of a
BPS bound together with the corresponding first-order BPS system for the
GPE. This discovery allows us to construct the first analytic configurations
with fractional vorticity. The reason why such a bound has not been found
before is that the topological charge on the right-hand side is not the
obvious one (which, in the present case, is the vorticity). This kind of
phenomenon (in which the right-hand side of the BPS bound is not
proportional to the most obvious topological charge) also appears in the low
energy limit of QCD: see \cite{BPSlast1,BPSlast2} and references therein. In
the present case, the topological charge is the sum of two terms. The first
one is determined by a topological density, which can be interpreted as the
vorticity dressed by the condensate profile. The second one has to do with
the \textquotedblleft skewness\textquotedblright\ of the configuration.

This paper is organized as follows. Section \ref{sec_GPE} introduces the
GPE, and a novel BPS bound is derived. In Section \ref{sec_TC}, the charges
that emerged from BPS bounds are interpreted. In Section \ref{sec_Solutions}%
, we show two kinds of analytical solutions from the BPS equations and
compute the associated topological charges. Section \ref{sec_AP_rel} shows a
general relation that the condensate amplitude and phase should satisfy.
Finally, Section \ref{sec_Conclution} is devoted to the conclusions. We will
provide the computational details in two appendices.

%%%%%%%%%%%%%%%%%%%%%%%%%%%%%%%%%%%%%%%%%%%%

\section{$2D$ Gross-Pitaevskii and a novel BPS bound}

\label{sec_GPE} %%%%%%%%%%%%%%%%%%%%%%%%%%%%%%%%%%%%%%%%%%%%

As is well known, the GPE in one spatial dimension is integrable, and
one-dimension vortices do not exist. Thus, we will consider the GPE in two
spatial dimensions \cite{R0,R01,R11}:
\begin{equation}  \label{GPE1}
\mathrm{i}\,\hbar\,\partial_{t}\Phi = -\frac{\hbar^{2}}{2M}\triangle \Phi
+g\left\vert \Phi \right\vert ^{2}\Phi +V\Phi -\mu \Phi \;,
\end{equation}
where
\begin{equation}  \label{amplitude-phase}
\Phi =\rho\, e^{\mathrm{i} \,S} \;,
\end{equation}
being $\mu$ the chemical potential, $g$ the coupling constant, $\Phi$ the
condensate wave function, $V$ is the external potential, $\rho$ is the
corresponding amplitude, and $S$ is the phase. One can easily obtain the
two-dimensional GPE with a confining harmonic potential in the $z$-direction
of the form $V_{conf}=\frac{\varpi }{2}z^{2}$.

In the present section, we will consider the most straightforward (but still
highly non-trivial) case of the stationary GPE in two spatial dimensions
with $V=0$ and $\mu =0$, i.e.,
\begin{equation}
-\frac{\hslash ^{2}}{2M}\triangle \Phi +g\left\vert \Phi \right\vert
^{2}\Phi =0\ .  \label{GPE1.1}
\end{equation}
The following sections will analyze strictly static solutions (without the
chemical potential term). The reason for this is that in the analysis of
turbulence in two dimensions, the energy term plays the central role, and,
in this term, the chemical potential is absent (see \cite{Shukla2013}, in
particular, Eqs. (2) and (3) of this reference). Indeed, one of the most
interesting applications of present formalism is the analysis of turbulence
in two dimensions (which we will discuss in a future publication). On the
other hand, the present formalism can be easily extended to the cases in
which extra linear terms are included in the GPE (see \cite{Canfora2025}).

Equation (\ref{GPE1.1}) can be derived from the following energy-functional
\begin{equation}
E=\int_{\Gamma }d^{2}x\left[ \frac{\hslash ^{2}}{2M}\left\vert
\overrightarrow{\nabla }\Phi \right\vert ^{2}+\frac{g}{2}\left\vert \Phi
\right\vert ^{4}\right] =\frac{\hslash ^{2}}{M}\int_{\Gamma }d^{2}x\left[
\frac{1}{2}\left\vert \overrightarrow{\nabla }\Phi \right\vert ^{2}+\frac{%
g_{eff}}{4}\left\vert \Phi \right\vert ^{4}\right] \;,  \label{GPE2}
\end{equation}
\begin{equation*}
g_{eff}=\frac{2Mg}{\hslash ^{2}}\;,
\end{equation*}%
where $\Gamma $ is the bounded region where the condensate is confined and $%
d^{2}x=dx\,dy$ ($x$ and $y$ being the spatial coordinates). The computations
in the following sections are simplified if, instead of considering $\Phi$
as one complex field, one considers $\Phi $ as two real scalar fields.
Ultimately, one can always return to the original notation in terms of $\Phi$%
. Thus, let us introduce the following notation
\begin{equation}
\phi _{1}=\rho \cos S\ ,\ \phi_{2}=\rho \sin S\ \Leftrightarrow \ \phi _{1}=%
\mbox{Re}  (\Phi )\ ,\ \phi_{2}=\mbox{Im}  (\Phi )\ ,\   \label{GPE2.1}
\end{equation}%
where $\mbox{Re} (\Phi)$ denotes the real part of $\Phi $ while $\mbox{Im}
(\Phi )$ denotes the imaginary part of $\Phi $. By using this notation, Eq. (%
\ref{GPE1.1}) becomes
\begin{eqnarray}
-\triangle \phi _{j}+g_{eff}\left( \overrightarrow{\phi }\cdot
\overrightarrow{\phi }\right) \phi _{j} &=&0\ ,\ \ j=1,2  \label{GPE2.2} \\
\left( \phi _{1}\right) ^{2}+\left( \phi _{2}\right) ^{2} &=&\overrightarrow{%
\phi }\cdot \overrightarrow{\phi }\;  \notag
\end{eqnarray}
while the energy becomes
\begin{equation}
E=\frac{\hslash ^{2}}{2M}\int_{\Gamma }d^{2}x\left[ \sum_{j=1}^{2}\left(
\overrightarrow{\nabla }\phi _{j}\right) ^{2}+\frac{g_{eff}}{2}\left(
\overrightarrow{\phi }\cdot \overrightarrow{\phi }\right) ^{2}\right] \;,
\label{GPE2.3}
\end{equation}
\begin{equation*}
\left( \overrightarrow{\nabla }\phi _{j}\right) ^{2}=\left( \partial
_{x}\phi _{j}\right) ^{2}+\left( \partial _{y}\phi _{j}\right) ^{2}\;.
\end{equation*}

It is a direct computation to show that the energy can be rewritten as
follows:
\begin{equation}
E=\frac{\hslash ^{2}}{2M}\int_{\Gamma }d^{2}x\left[ \left( \partial _{x}\phi
_{1}+\partial _{y}\phi _{2}+A\right) ^{2}+\left( \partial _{y}\phi
_{1}-\partial _{x}\phi _{2}+B\right) ^{2}\right] +Q_{1}+Q_{2}\;,
\label{BPS1}
\end{equation}%
where $A$, $B$ and $\kappa $ are defined as
\begin{equation}
A=\frac{\kappa }{\sqrt{2}}\left( \phi _{2}^{2}-\phi _{1}^{2}\right) \ \ ,\ \
B=\ -\sqrt{2}\,\kappa \phi _{1}\phi _{2}\;,\ \ \kappa ^{2}=g_{eff}>0\;.
\label{BPS2}
\end{equation}%
The topological charge is a combination of the following two boundary terms
\begin{equation}
Q_{1}=\frac{\hslash ^{2}}{M}\int_{\Gamma }d^{2}x\Lambda \quad \mbox{and}%
\quad Q_{2}=\frac{\hbar ^{2}\,\kappa }{\sqrt{2}\,M}\int_{\Gamma
}d^{2}x\left( \partial _{x}J^{x}+\partial _{y}J^{y}\right) \;,  \label{BPS3}
\end{equation}%
where we defined
\begin{eqnarray}
\Lambda &=&\partial _{x}(2\phi _{1}\partial _{y}\phi _{2})+\partial
_{y}(-2\phi _{1}\partial _{x}\phi _{2})=d\phi _{1}\wedge d\phi _{2}=d\rho
\wedge dS=d\omega \;,  \notag  \label{BPS4} \\
\omega &=&\rho dS\;,
\end{eqnarray}%
and
\begin{equation}
J^{x}=\phi _{1}\left( \frac{\phi _{1}^{2}}{3}-\phi _{2}^{2}\right) \;,\quad
J^{y}=\phi _{2}\left( \phi _{1}^{2}-\frac{\phi _{2}^{2}}{3}\right) \;.
\label{BPS5}
\end{equation}%
It is clear from Eqs. (\ref{BPS4}) and (\ref{BPS5}) that $Q_{1}$ and $Q_{2}$
are integrals of total derivatives.

Therefore, quite surprisingly, the following first order BPS system
\begin{eqnarray}
\partial _{x}\phi _{1}+\partial _{y}\phi _{2} &=&-A\;,  \label{BPSGPE1} \\
\partial _{y}\phi _{1}-\partial _{x}\phi _{2} &=&-B\;,  \label{BPSGPE2}
\end{eqnarray}%
actually implies the second-order GPE system in Eq. (\ref{GPE2.2}) and it is
a lower bound of the energy (\ref{BPS1}). This is a quite unexpected result
as it has been always implicitly assumed that the useful BPS bounds could
not be found in the case of the GPE in two or three spatial dimensions.

At this point, a natural question is in order:\textit{\ Can this approach be
somehow extended to the non-static case}? As has already been emphasized here above, in the present manuscript, we will focus on static configurations depending on two spatial coordinates, and, at first glance, the present approach is tied to the assumption of static configurations. Nevertheless, the remarkable similarity between the BPS
equations (\ref{BPSGPE1})-(\ref{BPSGPE2}) of the GPEs constructed here from
one side and the BPS equations for the magnetized BPS vortices in
Ginzburg-Landau theory for the other side strongly suggest the existence of
multi-vortices solutions in the present case as well. The details should be
polished, but the resemblance of the BPS equations (\ref{BPSGPE1})-(\ref%
{BPSGPE2}) with a ``zero-curvature condition'' for a complex connection
provides us with concrete hopes to achieve the goal of constructing
multi-vortices solutions. Indeed, a zero curvature condition for a
non-Abelian connection $\Xi $ reads
\begin{equation*}
d\Xi +\Xi \wedge \Xi =0\ ,
\end{equation*}%
where $\Xi $\ is a one-form taking values in the Lie algebra of a suitable
Lie group. Hence, one can see that the linear terms contain derivatives of
the connection $\Xi $ while the non-linear terms are quadratic in $\Xi $ and
without derivatives. This is what happens for the BPS equations in
Eqs. (\ref{BPSGPE1})-(\ref{BPSGPE2}) (as both $A$ and $B$ in Eq. (\ref{BPS2}%
) are quadratic in the fields). Thus, assuming that multi-vortices solutions
can be constructed using the standard techniques available in the presence
of a zero-curvature representation (see, for instance \cite%
{Hitchin1999,Babelon2003,Jurdjevic2016}), one can introduce the
time-dependence by promoting the moduli of the solutions (which represent
the positions of the vortices) to time-dependent functions. Then, the
dynamics of the vortices will be determined by the geodesics on the
corresponding moduli space metric (see \cite{modulia,modulib,moduli2,moduli3}
and references therein). We will come back on this interesting issue in a
future publication.

The GPE, and also the lower energy bound, can be obtained by a slight
generalization of $A$ and $B$ given in (\ref{BPS2}). In fact, if
\begin{eqnarray}
A &=&\frac{a_{1}}{2}\left( \phi _{2}^{2}-\phi _{1}^{2}\right) -a_{2}\phi
_{1}\,\phi _{2}  \notag  \label{A_B_mod} \\
B &=&-\frac{a_{2}}{2}\left( \phi _{2}^{2}-\phi _{1}^{2}\right) -a_{1}\,\phi
_{1}\phi _{2}\;,
\end{eqnarray}%
along with condition $a_{1}=\frac{\kappa }{\sqrt{2}}\,\cos \,\gamma $, and $%
a_{2}=\frac{\kappa }{\sqrt{2}}\,\sin \,\gamma $, satisfies the GPE for an
arbitrary real parameter $\gamma $ (see Appendix \ref%
{appendix_solutions_without_rotation} for details). In such a case, $Q_{1}$
does not change, while, by using some trigonometric identities
\begin{align}
J^{x}& =\cos \,\gamma \,\phi _{1}\,\left( \frac{\phi _{1}^{2}}{3}-\phi
_{2}^{2}\right) +\sin \,\gamma \,\phi _{2}\left( \phi _{1}^{2}-\frac{\phi
_{2}^{2}}{3}\right) =\frac{\rho ^{3}}{3}\,\cos (3S-\gamma )\;,  \label{BPS6}
\\
\quad J^{y}=& \cos \,\gamma \,\phi _{2}\,\left( \phi _{1}^{2}-\frac{\phi
_{2}^{2}}{3}\right) +\sin \,\gamma \,\phi _{1}\left( \phi _{2}^{2}-\frac{%
\phi _{1}^{2}}{3}\right) =\frac{\rho ^{3}}{3}\,\sin (3S-\gamma )\;.
\end{align}

%\rc{\textbf{XX check aca` arriba: he cambiado algo en la formula, habia un typo XX}}

Another natural question is: Can the BPS system be modified so that
potential-like linear terms in the condensate field appear in the
corresponding second-order GPE? The answer is positive. One example of the
modified expressions for $A$ and $B$, which achieve this goal is:
\begin{equation}
A=\frac{\kappa }{\sqrt{2}}\left( \phi _{2}^{2}-\phi _{1}^{2}\right) +f\phi
_{1}\ \ ,\ \ B=\ -\sqrt{2}\,\kappa \phi _{1}\phi _{2}+f\phi _{2}\;.
\label{modpot}
\end{equation}%
In a future publication, we will discuss how it is possible to choose the
function $f=f(x,y)$ to get interesting potential terms in the corresponding
GPE (describing, for instance, the effects of the chemical potential or a
central potential).

One can rewrite the first-order system (\ref{BPSGPE1})--(\ref{BPSGPE2}) by
using the \textquotedblleft amplitude-phase\textquotedblright\
representation for the BEC wave function in Eq. (\ref{GPE2.1}); hence, in
terms of the amplitude $\rho $ and the phase $S$, the first-order BPS system
in Eqs. (\ref{BPSGPE1}) and (\ref{BPSGPE2}) reduces to (see details in
Appendix \ref{appendix_solutions_without_rotation})%
\begin{eqnarray*}
\frac{\partial \,\rho }{\partial r}+\frac{\rho }{r}\frac{\partial \,S}{%
\partial \theta } &=&\frac{\kappa }{\sqrt{2}}\,\rho ^{2}\,\cos \,(\theta
-3S+\gamma )\;, \\
\rho \frac{\partial \,S}{\partial r}-\frac{1}{r}\frac{\partial \,\rho }{%
\partial \theta } &=&\frac{\kappa }{\sqrt{2}}\,\rho ^{2}\,\sin \,(\theta
-3S+\gamma )\;, \\
\rho &=&\rho (r,\theta )\ ,\ \ S=S(r,\theta )\ .
\end{eqnarray*}%
The above form is very convenient for the hydrodynamic interpretation. In
particular, let us introduce the variable $u=\frac{1}{\rho }$ together with
the superfluid velocity $\overrightarrow{V}=\overrightarrow{\nabla }S$. In
terms of $u$ and $\overrightarrow{V}$, the above BPS system can be written
as
\begin{eqnarray*}
-\frac{\partial \,u}{\partial r}+\frac{u}{r}\,V_{\theta } &=&\frac{\kappa }{%
\sqrt{2}}\,\cos \,(\theta -3S+\gamma )\;, \\
u\,V_{r}+\frac{1}{r}\frac{\partial \,u}{\partial \theta } &=&\frac{\kappa }{%
\sqrt{2}}\,\sin \,(\theta -3S+\gamma )\;, \\
V_{\theta } &=&\frac{\partial \,S}{\partial \theta }\ ,\ V_{r}=\frac{%
\partial \,S}{\partial r}\ .
\end{eqnarray*}%
In this form, it is easy to deduce many interesting properties from the BPS
system. First of all, when $\left\vert \theta -3S+\gamma \right\vert $ is
small, one can see that a $\theta$-dependence of the amplitude $\rho $ of
the wave function (which is the superfluid density) induces a radial
component of the superfluid velocity. Moreover, in the regions where $\left(
\theta -3S+\gamma \right) $ is constant, the BPS equations became algebraic
for $\rho $ and the velocity $\overrightarrow{V}$ where $r$ and the coupling
$\kappa $\ appear as parameters. One can see that (unless one restricts the
attention to the most symmetric cases) regions where $V_{r}$ is not
vanishing appear. In these regions, $V_{r}$ is tied to $\frac{\partial
\,\rho }{\partial \theta }$. Thus, the availability of a BPS system for the
GPE allows us to derive relevant hydrodynamic properties of the superfluid
in a very natural way.

\subsection{On possible generalizations of the present approach}

A natural criticism of the present approach is that, at first glance, it is
closely related to the form of the quartic potential characterizing the
usual GPE. In fact, this is not the case. Let us consider, for instance, a
complex scalar field with a non-linear interaction of order six (which also
is relevant in the theory of superfluidity as well as in relativistic field
theories: see, for instance, \cite{CS1,CS2,CS3,CS4a,CS4b}). The energy of
static field configurations read in this case
\begin{equation}
E=\frac{\hslash ^{2}}{2M}\int_{\Omega }d^{2}x\left[ \sum_{j=1}^{2}\left(
\overrightarrow{\nabla }\Phi _{j}\right) ^{2}+\frac{g_{eff}}{2}\left(
\overrightarrow{\Phi }\cdot \overrightarrow{\Phi }\right) ^{3}\right] \ ,
\label{GPE63}
\end{equation}%
\begin{equation*}
\left( \overrightarrow{\nabla }\Phi _{j}\right) ^{2}=\left( \partial
_{x}\Phi _{j}\right) ^{2}+\left( \partial _{y}\Phi _{j}\right) ^{2}\ ,
\end{equation*}%
\begin{equation*}
\Phi _{1}=\rho \cos S\ ,\ \Phi _{2}=\rho \sin S\ \Leftrightarrow \ \Phi _{1}=%
\mbox{Re} \Psi \ ,\ \Phi _{2}=\mbox{Im} \Psi \ ,\ \ \Psi =\rho \exp iS\ ,\
\rho \geq 0\ .\
\end{equation*}

On the other hand, the field equations (the generalized GPE) read
\begin{eqnarray}
-\triangle \Phi _{j}+g_{eff}\left( \overrightarrow{\Phi }\cdot
\overrightarrow{\Phi }\right) ^{2}\Phi _{j} &=&0\ ,\ \ j=1,2  \label{GPE61}
\\
\left( \Phi _{1}\right) ^{2}+\left( \Phi _{2}\right) ^{2} &=&\overrightarrow{%
\Phi }\cdot \overrightarrow{\Phi }\ ,\ \ g_{eff}=\kappa \ .  \notag
\end{eqnarray}%
With a technique similar to the one used in the previous section, one can
rewrite the energy as follows
\begin{equation}
E=\frac{\hslash ^{2}}{2M}\int_{\Gamma }d^{2}x\left[ \left( \partial _{x}\Phi
_{1}+\partial _{y}\Phi _{2}-\widehat{\Theta }_{1}\right) ^{2}+\left(
\partial _{y}\Phi _{1}-\partial _{x}\Phi _{2}-\widehat{\Theta }_{2}\right)
^{2}\right] +\widehat{Q}_{1}+\widehat{Q}_{2}\;,  \label{energysix1}
\end{equation}%
where $\widehat{\Theta }_{1}$ and $\widehat{\Theta }_{1}$ are defined as
\begin{equation}
\widehat{\Theta }_{1}=\frac{\kappa }{\sqrt{3}}\,(\Phi _{1}^{3}-3\Phi
_{1}\Phi _{2}^{2})\ \ ,\ \ \widehat{\Theta }_{2}=\frac{\kappa }{\sqrt{3}}%
\,\left( 3\Phi _{1}^{2}\,\Phi _{2}-\Phi _{2}^{3}\right) \;.
\label{energysix2}
\end{equation}

Thus, quite remarkably, also in this theory, one arrives at the following
novel BPS system
\begin{eqnarray}
\frac{\partial \,\Phi _{1}}{\partial x}+\frac{\partial \,\Phi _{2}}{\partial
y} &=&\frac{\kappa }{\sqrt{3}}\,(\Phi _{1}^{3}-3\Phi _{1}\Phi _{2}^{2})\;,
\label{BPS61} \\
\frac{\partial \,\Phi _{1}}{\partial y}-\frac{\partial \,\Phi _{2}}{\partial
x} &=&\frac{\kappa }{\sqrt{3}}\,\left( 3\Phi _{1}^{2}\,\Phi _{2}-\Phi
_{2}^{3}\right) \;,  \label{BPS62}
\end{eqnarray}%
whose solutions are automatic solutions of the field equations in Eq. (\ref%
{GPE61}). The interest in this case is that it explicitly shows that the
fraction appearing in the angular dependence of the phase $S(\theta )$ of
the condensate of the simplest non-trivial solution of the above BPS
equations (which determines the fractional vorticity) \textit{does depend on
the self-interaction potential}. Such solution of the BPS equations (\ref%
{BPS61}) and (\ref{BPS62}) in cylindrical coordinates (namely, within the
metric $ds^{2}=dr^{2}+r^{2}d\theta ^{2}$) reads
\begin{eqnarray}
\rho (r) &=&(A\,r^{1/2}-\frac{4}{\sqrt{3}}\,\kappa \,r)^{-\frac{1}{2}}\;,
\label{sol_n_3} \\
S(\theta ) &=&\frac{\theta }{4}+S_{0}\;,  \notag
\end{eqnarray}%
where $A$ and $S_{0}$ are two integration constants (so that the
corresponding fraction is $1/4$ in the case of the $\left( \overrightarrow{%
\Phi }\cdot \overrightarrow{\Phi }\right) ^{3}$\ interaction).

In this case too the topological charge is a combination of the following
two boundary terms:
\begin{equation}
\widehat{Q}_{1}=\frac{\hslash ^{2}}{M}\int_{\Gamma }d^{2}x\Lambda \quad %
\mbox{and}\quad \widehat{Q}_{2}=\frac{\hbar ^{2}}{2M}\int_{\Gamma
}d^{2}x\left( \partial _{x}J_{(6)}^{x}+\partial _{y}J_{(6)}^{y}\right) \;,
\label{top5}
\end{equation}%
where we defined
\begin{eqnarray}
\Lambda  &=&\partial _{x}(2\Phi _{1}\partial _{y}\Phi _{2})+\partial
_{y}(-2\Phi _{1}\partial _{x}\Phi _{2})=d\Phi _{1}\wedge d\Phi _{2}=d\rho
\wedge dS=d\omega \;,  \notag \\
\omega  &=&\rho dS\;,
\end{eqnarray}%
and
\begin{equation}
J_{(6)}^{x}=\frac{\,\kappa }{2\,\sqrt{3}}\,\left( \Phi _{1}^{4}+\Phi
_{2}^{4}-6\,\Phi _{1}^{2}\,\Phi _{2}^{2}\right) \;,\quad J_{(6)}^{y}=\frac{%
\kappa }{2\,\sqrt{3}}\,\left( 4\Phi _{1}^{3}\,\Phi _{2}-4\Phi _{2}^{3}\,\Phi
_{1}\right) \;,  \label{top6}
\end{equation}%
where the notation $J_{(6)}^{x}$ and $J_{(6)}^{y}$\ is to remind that this
topological current in Eqs. (\ref{top5}) and (\ref{top6}) refers to the
interaction potential of order six.

To the best of the author's knowledge, this is the first example of a BPS
bound in the theory of a complex scalar field with non-linear interactions
of order six for field configurations depending on two spatial coordinates.
Moreover, the present results can be further extended to more general forms
of self-interactions potentials (see \cite{Canfora2025}).

\section{Interpretation of the topological charge}

\label{sec_TC}

The term $Q_{1}$ is related, as expected, to the vorticity. Although it is
not exactly the number of vortices, it is proportional to it when the
amplitude $\rho $ is equal to a constant. This is very similar to what
happens in chiral perturbation theory, where useful BPS bounds can be
achieved provided the \textquotedblleft obvious\textquotedblright\
topological densities are dressed by the corresponding hadronic profiles
(see \cite{BPSlast1,BPSlast2} and references therein). In a sense, one can
say that $Q_{1}$ represents the \textquotedblleft natural\textquotedblright\
topological charge in the present context of vortices in superfluids.
Indeed, we can write $Q_{1}$ as
\begin{equation}
Q_{1}=\frac{\hslash ^{2}}{M}\int_{\Gamma }d^{2}x\Lambda =\frac{\hslash ^{2}}{%
M}\int_{\partial \Gamma }\overrightarrow{\omega }\cdot d\overrightarrow{l}\;,
\label{charge1}
\end{equation}%
where $\Gamma $ is a given region (usually, but not always, a disk of radius
$R$), $d\overrightarrow{l}$ is the line element tangent to $\partial \Gamma $%
. Clearly, $Q_{1}$ is related to the vorticity as, for instance, if we give
boundary conditions at spatial infinity (so the radius $R$ of the disk $%
\Gamma $ approaches to infinity) such that the amplitude $\rho $ goes to $1$%
, then $Q_{1}$ becomes exactly the vorticity (times $\frac{\hslash ^{2}}{M}$%
):
\begin{equation*}
\rho \underset{r\rightarrow \infty }{\rightarrow }1\ \ \mbox{and}\ \ \
\partial \Gamma =\mbox{``the\ circle\ at\ infinity''}\ \Rightarrow Q_{1}=%
\frac{\hslash ^{2}}{M}\oint_{\partial \Gamma }\overrightarrow{\nabla }S\cdot
d\overrightarrow{l}=\frac{\hslash ^{2}}{M}(\mbox{Vorticity})\;.
\end{equation*}%
The same is true if one requires that $\rho =1$ on $\partial \Gamma $\ even
when $\partial \Gamma $ is not the circle at infinity. However, in general, $%
Q_{1}$ is the vorticity or circulation dressed by the profile $\rho $ of the
condensate as it happens in non-linear sigma models and chiral perturbation
theory \cite{BPSlast1,BPSlast2}.

On the other hand, $Q_{2}$ is not directly related to the vorticity, and it
can be written as
\begin{equation}
Q_{2}=\frac{\hbar ^{2}\,\kappa}{\sqrt{2}\,M}\int_{\Gamma }d^{2}x\,\left(
\partial_{x}J^{x}+\partial _{y}J^{y}\right) =\frac{\hbar^{2}\,\kappa}{\sqrt{2%
}\,M}\oint_{\partial\Gamma }\hat{n}\cdot \overrightarrow{J}\,ds\;,
\label{charge2}
\end{equation}
where $\overrightarrow{n}$ is the unit outer normal to $\partial \Gamma $.

A possible interpretation of $Q_{2}$ (the topological current being defined
in Eq. (\ref{BPS5})) is that it measures, in a sense, the \textquotedblleft
skewness\textquotedblright\ of the configuration. Indeed, from Eq. (\ref%
{BPS5}), one can see that
\begin{equation*}
J^{x}=\phi _{1}\left( \frac{\phi _{1}^{2}}{3}-\phi _{2}^{2}\right) =\frac{1}{%
3}\mbox{Re}\left[ \left( \phi _{1}+i\phi _{2}\right) ^{3}\right] \;,\quad
J^{y}=\phi _{2}\left( \phi _{1}^{2}-\frac{\phi _{2}^{2}}{3}\right) =\frac{1}{%
3}\mbox{Im}\left[ \left( \phi _{1}+i\phi _{2}\right) ^{3}\right] \ .
\end{equation*}%
Thus, if the vector ($J^{x}$, $J^{y}$) behaves as a magnetic field (which is
divergence-free) within $\Gamma $, then $Q_{2}$ will vanish. If, instead,
the vector ($J^{x}$, $J^{y}$) behaves as an electric field with some source
within $\Gamma $, $Q_{2}$ will not vanish. Close to the source (and assuming
that such a source is point-like), the vector ($J^{x}$, $J^{y}$) points
radially outward. Hence, at least close to the source, $J^{x}$ and $J^{y}$
will be the same size and (in a small neighborhood of the source) the vector
($J^{x}$, $J^{y}$) will be invariant under rigid rotations. Consequently,
when $Q_{2}$ does not vanish, $\mbox{Re}\left[ \left( \phi _{1}+i\phi
_{2}\right) ^{3}\right] $ and $\mbox{Im}\left[ \left( \phi _{1}+i\phi
_{2}\right) ^{3}\right] $ (at least close to the source) can be rotated one
into the other. On the other hand, in the \textquotedblleft magnetic
case\textquotedblright\ (in which $\partial _{x}J^{x}+\partial _{y}J^{y}$\
vanishes identically) $\mbox{Re}\left[ \left( \phi _{1}+i\phi _{2}\right)
^{3}\right] $ and $\mbox{Im}\left[ \left( \phi _{1}+i\phi _{2}\right) ^{3}%
\right] $ play a very different role. Indeed, when $\Gamma $ is small
enough, the tangent lines to the vector field ($J^{x}$, $J^{y}$) are close
to be parallel lines. Therefore, when $\Gamma $ is small enough, with a
rigid rotation one can eliminate one of the two components of the vector ($%
J^{x}$, $J^{y}$). We can resume the above arguments by saying that $Q_{2}$
(especially when the domain $\Gamma $ is small enough) is a measures of how
similar is the behavior of $\mbox{Re}\left[ \left( \phi _{1}+i\phi
_{2}\right) ^{3}\right] $ and $\mbox{Im}\left[ \left( \phi _{1}+i\phi
_{2}\right) ^{3}\right] $. The same reasoning also holds in the case of the
non-linear self-interaction of order six discussed in the previous section.

\section{Analytic solutions and fractional vorticity from the BPS equations}

\label{sec_Solutions}

In this section, we will present two types of GPE system solutions (\ref%
{BPSGPE1})--(\ref{BPSGPE2}) and some of their main features.

\subsection{$1/3-$ fractional vorticity configuration}

By using the BPS equations (\ref{BPSGPE1}) and (\ref{BPSGPE2}), we can
construct the first analytic solution for the GPE (see Appendix \ref%
{appendix_solutions_without_rotation} for details) with fractional
vorticity:
\begin{eqnarray}
\phi _{1}(r,\theta ) &=&\frac{2\sqrt{2}\,\cos \,\left( \frac{\theta }{3}%
+\theta _{0}\right) }{2\sqrt{2}\,A\,r^{\frac{1}{3}}-\,3\kappa \,r}\;,  \notag
\label{solGPE1} \\
\phi _{2}(r,\theta ) &=&\frac{2\sqrt{2}\,\sin \,\left( \frac{\theta }{3}%
+\theta _{0}\right) }{2\sqrt{2}\,A\,r^{\frac{1}{3}}-\,3\kappa \,r}\;.
\end{eqnarray}%
where $A$ and $\theta _{0}$ are real integration constants.

As far as the ``physical interpretation of the solution'' is concerned, it
is pretty clear that such solutions cannot appear in isolation in a
two-dimensional plane because these fractional configurations do not respect
the (2$\pi$)-period of theta (on the other hand, these configurations can be
well-defined on cones with the same deficit angle). Instead, the interest of
such solutions lies in the fact that their existence shows the exciting
possibility of having fractional vorticity without many extra ingredients.

A few comments are in order. First of all, the solution is defined on an
annulus of inner radius $R_{1}$, and outer radius $R_{2}$ ($R_{1}<R_{2}$)
provided that the integration constant $A$ is chosen in such a way that both
$\phi _{i}(r,\theta )$ defined here above are regular for any $r\in \left[
R_{1},R_{2}\right] $ (this is always possible). It is worth emphasizing that
in effective field theories (such as the mean-field theory of superfluidity
described by the GPE), it is natural to introduce an ultraviolet cutoff $a$
(which represents the radius of the atoms of the superfluid: see \cite{R0}
and references therein). Thus, in principle, one could take $R_{1}$ of order
$a$. Secondly, the above analytic solution of the GPE possesses the
following peculiar feature: namely, the phase $S$ does not cover ``the full
circle'' since $S=\frac{\theta }{3}+\gamma $ ($\gamma $ being a constant) so
that when $\theta $ goes from $0$ to $2\pi $, the variation $\Delta S$ of
the phase is $\frac{2\pi }{3}$. Thus, the BPS equations (\ref{BPSGPE1}) and (%
\ref{BPSGPE2}) for an annulus-like configuration force the vorticity to be $%
1/3$. The formation of a configuration with fractional vorticity is a quite
remarkable phenomenon that \emph{only requires} the BPS structure disclosed
in the present manuscript in the GPE and no further ingredients.

The appearance of fractional vortices and, in general, of configurations
with fractional topological charges has been discussed in condensed matter
physics (see \cite{FracVort1,FracVort2,FracVort3,FracVort4} and references
therein), in high energy physics (see \cite%
{FracVort5,FracVort6,FracVort7,FracVort8} and references therein) since long
time ago, and even in laser fields (see for instance \cite{FracVort9} and
references therein). It is fair to say (as it is evident from the reference
mentioned above) that the simplest ingredients are not enough to achieve
configurations with fractional topological charges. Often, either extra
fields/interactions are included in the Lagrangian, or fractional
topological excitations are not solutions of the classical equations of
motion but arise as quantum solutions of the effective Lagrangian. Quite
surprisingly, in the present context, \emph{the possibility to have
configurations with fractional vorticity arises simply from the BPS system
of the GPE, the factor} $1/3$ \emph{being related to the cubic non-linear
term characterizing the GPE itself}. In other words, no extra ingredient is
needed in the GPE case, the fraction being fixed by the cubic non-linear
interaction appearing in the GPE.

Coming back to the notation $\phi =\rho \,e^{\mathrm{i}\,S}$, and the polar
coordinates $(r,\theta )$, one can show that, when $S=\frac{\theta }{3}$,
the second-order GPE reduces to
\begin{equation}
\rho ^{\prime \prime }+\frac{\rho ^{\prime }}{r}-\frac{\rho }{9r^{2}}%
-g_{eff}\,\rho ^{3}=0\;.  \label{GPEspherical}
\end{equation}%
The factor $-\frac{1}{9}$ comes from $S=\frac{\theta }{3}+\gamma$. It is a
direct computation to show that the solution of the BPS equation (\ref%
{solGPE1}) is also a solution of the GPE equation in Eq. (\ref{GPEspherical}%
), whose amplitude profile is
\begin{equation}
\rho (r)=\frac{2\sqrt{2}}{2\sqrt{2}\,A\,r^{\frac{1}{3}}-3\kappa \,r}\;,
\label{amplitude_solGPE1}
\end{equation}%
being $A$ an integration constant to be fixed to have $\rho (r)$ positive in
the region of interest. To the best of the authors' knowledge, it is worth
emphasizing that (\ref{solGPE1}) is the first analytic solution of the GPE
with non-vanishing fractional vorticity and a non-trivial radial profile.
Let us stress the important fact that both $\pm \kappa$ correspond to the
same GPE (\ref{GPEspherical}), as $g_{eff}$ is proportional to $\kappa ^{2}$.

\begin{figure}[h!]
\centering
\includegraphics[scale=0.65]{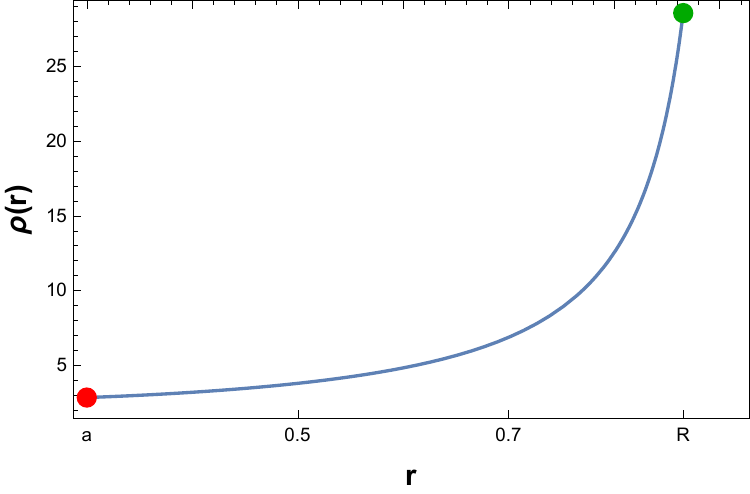}
\caption{The profile $\protect\rho $ vs. the radial coordinate $r$ for the
fractional vorticity solution found with BPS bounds. We took $\protect\kappa %
=1$ and $A=1$, and internal radius $a=0.3$ (red dot) while the external one
is $R\approx0.91$ (dark-green dot) in arbitrary units.}
\label{fig_first_sol}
\end{figure}

\begin{figure}[h!]
\begin{subfigure}[b]{0.45\textwidth}
\includegraphics[width=\textwidth]{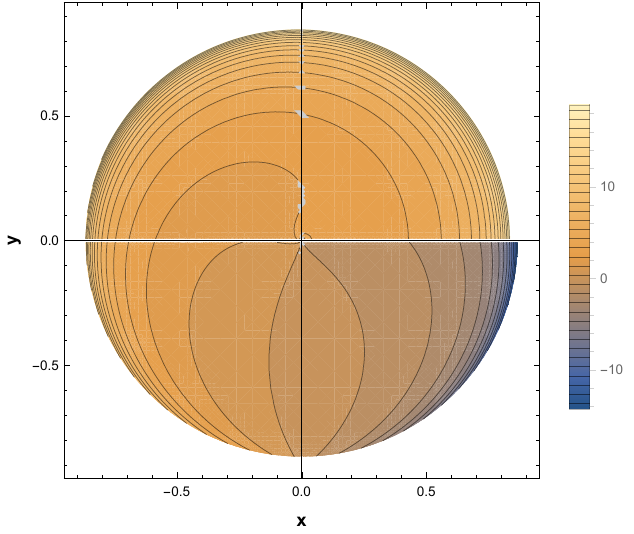}
\caption{Level curves for $\phi_{1}$ of solution (\ref{solGPE1}). We see a branch cut in the positive $x$-axis due to the $\frac{\theta}{3}$ angular dependence of $\phi_{1}$. }
\end{subfigure}
\hfill
\begin{subfigure}[b]{0.45\textwidth}
\includegraphics[width=\textwidth]{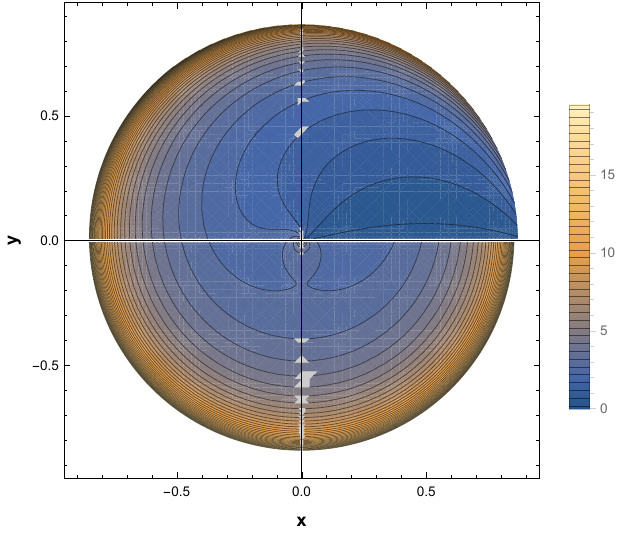}
\caption{Level curves for $\phi_{2}$ of solution (\ref{solGPE1}). We see a branch cut in the positive $x$-axis due to the $\frac{\theta}{3}$ angular dependence of $\phi_{2}$.}
\end{subfigure}
\caption{Level curves for solution (\protect\ref{solGPE1}). We took $\protect%
\kappa =1$ and $A=1$, and internal radius $a=0.3$ while the external one is $%
R\approx0.91$ in arbitrary units.}
\label{fig_level_first_sol}
\end{figure}

In Figure \ref{fig_first_sol}, it is shown the profile of $\rho $ for
solution (\ref{solGPE1}) for $A=1$ and $\kappa =-1$ (note that both $\kappa $
and $A$ has dimensions of mass in natural units). We can see that such a
profile does not depend on $\theta _{0}$ parameter.
Regarding the fields in Eq. (\ref{solGPE1}), Figure \ref{fig_level_first_sol}
shows the level curves for both of them. There is a branch cut (which can be
chosen on the positive $x$-axis), and this is expected from the argument $%
\frac{\theta}{3}$ in (\ref{solGPE1}). Hence, it is natural to represent the
solution on a Riemann surface as it happens in the cases of defects with
angular excess living on surfaces with negative intrinsic curvature (see
\cite{Kleinert} and references therein). These kinds of defects are also
relevant to use graphene as an effective playground to test important
features of quantum field theories in curved spacetimes \cite%
{Iorio2013,IorioReview,Iorio2018}. We hope to come back to this issue in a
future publication.
\begin{comment}In the following section we will discuss how to modify the BPS equation in such a way to obtain both the phase $S$ covering "the full circle" and a regular
amplitude $\rho $.

\textbf{XX evaluar }$Q_{1}$\textbf{\ y }$Q_{2}$\textbf{XX}

\textbf{XXX comentar que GPE no fija el coeficiente de }$\theta $\textbf{\
en la fase pero la BPS si lo fijan XXX}
\end{comment}

For the solution defined above, the two boundary terms $Q_{1}$ and $Q_{2}$
contributing to the topological charge can be computed explicitly. Taking
into account Eq. (\ref{charge1}) for $S_{0}=0$, in an annulus one gets
\begin{eqnarray}  \label{Q1_first_sol}
Q_{1} &=&-\frac{\hbar ^{2}}{M}\int_{0}^{2\pi }\,\rho ^{2}\,\cos
^{2}S\,\partial _{\theta }S\,d\theta  \notag \\
&=&-\frac{(8\pi + 3\sqrt{3})\,\hbar ^{2}}{3M(2\,\sqrt{2}\,A\,R^{\frac{1}{3}%
}-\,3\kappa \,R)^{2}} + \frac{(8\pi + 3\sqrt{3})\,\hbar ^{2}}{%
3M(2^{1/3}A\,a^{\frac{1}{3}}-\,3\kappa \,a)^{2}}\;,
\end{eqnarray}
where $R$ is the outer radius, and $a$ is the inner (which could be taken as
the size of the condensed atoms of interest). On the other hand, the
vorticity is $1/3$ as anticipated.

As far as $Q_{2}$ is concerned, one gets at the same annulus
\begin{comment}
The $Q_{2}$ charge has also a compact form. Indeed, taking into account that
\begin{eqnarray*}
\partial_{x}J^{x}+\partial _{y}J^{y} &=& (\phi_{1}^{2}-\phi_{2}^{2})\,(\partial_{x}\phi_{1}+\partial_{y}\phi_{2}) + 2\phi_{1}\,\phi_{2}\,(\partial_{y}\phi_{1}-\partial_{x}\phi_{2}) = \frac{\kappa^{2}}{\sqrt{2}}(\phi_{1}^{2} + \phi_{2}^{2})^{2} \;,\ \
\end{eqnarray*}
where in the last equality we used the BPS equations (\ref{BPSGPE1}) and (\ref{BPSGPE2}).
Then,
\end{comment}%
\begin{equation}
Q_{2}=\frac{\hslash ^{2}\,\kappa }{\sqrt{2}\,M}\oint_{\partial \Gamma }\hat{n%
}\cdot \overrightarrow{J}\,ds\;,  \label{charge2}
\end{equation}%
where
\begin{eqnarray*}
J^{x} &=&\frac{\rho ^{3}}{3}\,\cos (3S-\gamma )\,, \\
J^{y} &=&\frac{\rho ^{3}}{3}\,\sin (3S-\gamma )\;.
\end{eqnarray*}%
Hence, we have
\begin{eqnarray*}
\hat{n}\cdot \overrightarrow{J} &=&\cos \theta J^{x}+\sin \theta J^{y}=\frac{%
\rho ^{3}}{3}\left( \cos \theta \,\cos (3S-\gamma )+\sin \theta \,\sin
(3S-\gamma )\right) \\
&=&\frac{\rho ^{3}}{3}\,\cos (3S-\gamma -\theta )=\frac{\rho ^{3}}{3}\;.
\end{eqnarray*}%
Therefore, we obtain
\begin{eqnarray}
Q_{2} &=&\frac{32\,\pi \,\hbar ^{2}\,\kappa }{{3M}(2\,\sqrt{2}A-3\kappa
\,R^{2/3})^{3}}\,  \notag  \label{Q2_first_sol} \\
&-&\frac{32\,\pi \,\hbar ^{2}\,\kappa }{{3M}(2\,\sqrt{2}A-3\kappa
\,a^{2/3})^{3}}\;.
\end{eqnarray}%
One can see that $Q_{2}$ is still finite when $a\rightarrow 0$ provided $%
A\neq 0$. Moreover, when $a\rightarrow 0$ and $R\rightarrow +\infty $, $%
Q_{2} $ approaches to
\begin{equation}  \label{Q2_first_solution}
Q_{2}=-\frac{16\,\pi \,\hbar ^{2}\,\kappa }{3\sqrt{2}\,M\,A^{3}}\;.
\end{equation}
Therefore, $Q_{2}$ is inversely proportional to the cube of $A$ if the
annulus is extended to the whole $xy$ plane.

\subsection{Domain-wall type solutions}

Another analytical solution of the GPE can be easily obtained by solving the
BPS equations (details are in Appendix \ref%
{appendix_solutions_without_rotation}). It takes a simpler form in Cartesian
coordinates:
\begin{eqnarray}
\phi _{1}(x,y) &=&\frac{\sqrt{2}\,\cos \,S_{0}}{\sqrt{2}\,A-x\,\kappa
\,\,\cos \,(3S_{0}-\gamma )-y\,\kappa \,\sin \,(3S_{0}-\gamma ))}\;,
\label{solGPE2} \\
\phi _{2}(x,y) &=&\frac{\sqrt{2}\,\sin \,S_{0}}{\sqrt{2}\,A-x\,\kappa
\,\,\cos \,(3S_{0}-\gamma )-y\,\kappa \,\sin \,(3S_{0}-\gamma ))}\;.
\end{eqnarray}%
where $A$, $S_{0}$ and $\gamma $ are integration constants. This solution
could represent two disjoint regions separated by a wall. It is easy to see
that the amplitude diverges when
\begin{equation*}
\sqrt{2}\,A-x\,\kappa \,\,\cos \,(3S_{0}-\gamma )-y\,\kappa \,\sin
\,(3S_{0}-\gamma ))=0\;.
\end{equation*}%
The divergence appears on a line of slope $-\cot \,(3S_{0}-\gamma )$ and $y$%
intercept at $\frac{\sqrt{2}A}{\kappa \,\sin (3S_{0}-\gamma )}$. Thus, one
can assume that the position of the wall is precisely at the location where
the amplitude diverges. As usual, a natural cut-off is $a$ (the size of the
atoms of the superfluid).

The amplitude profile $\rho $ must be non-negative in order to represent a
sensible solution. For instance, we take $\kappa $ positive, $A=0$ and $%
3S_{0}-\gamma =\frac{\pi }{4}$, then the physical region is $x+y<0$. On the
other hand, if we take negative $\kappa $, the physical region is $x+y>0$.
This piecewise $\rho $ is well-defined except in the line $x+y=0$. In Figure %
\ref{fig_second_sol} it is shown the amplitude profile $\rho $ for this kind
of solution. It is worth mentioning that this analytic solution has the same
behavior of the numerical soliton solutions of the GPE equation in \cite%
{Geiser2019}.

\begin{figure}[h]
\label{fig_second_sol}
\begin{subfigure}[b]{0.45\textwidth}
\includegraphics[width=\textwidth]{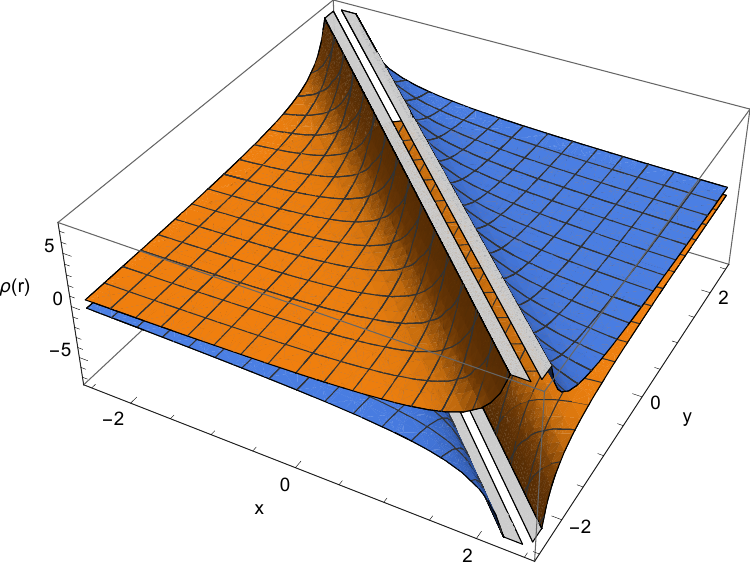}
\caption{ The minus sign in the denominator of (\ref{solGPE2}) is represented in blue, while the plus sign is in orange.}
\end{subfigure}
\hfill
\begin{subfigure}[b]{0.45\textwidth}
\includegraphics[width=\textwidth]{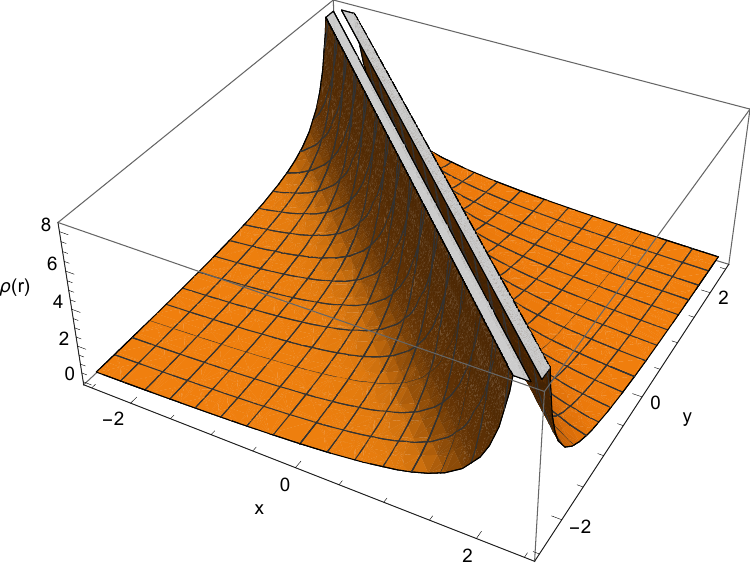}
\caption{The profile $\rho$ taken as a disjoint union of the plus and minus signs to have a non-negative value in the entire plane except in the line $x+y=0$.}
\end{subfigure}
\caption{The profile $\protect\rho $ vs. the radial coordinate $r$ for the
second solution found with BPS bounds. We took $A=0$, $\protect\kappa =1$,
and $3S_{0}+\protect\gamma =\frac{\protect\pi }{4}$. The profile was taken
such that the points closer than $a\approx 0.05$ to the line $x+y=0$ were
excluded.}
\label{fig_second_sol}
\end{figure}

This solution has vanishing vorticity as the phase is constant. In Figure %
\ref{fig_vector} it is shown the vector plot for $\vec{J}$. There, it is
evident that $\vec{J}$ is always perpendicular to the position of the wall
(namely, the line $\sqrt{2}\,A-x\,\kappa\,\,\cos\,(3S_{0}-\gamma
)-y\,\kappa\,\sin \,(3S_{0}-\gamma ))=0$). Therefore, if we take a region
excluding this line of thickness $a$ (in this case $\approx 0.05$), both $%
Q_{1}$ and $Q_{2}$ vanish:
\begin{equation}
Q_{1}=Q_{2}=0\;.  \label{Q1_and_Q2_second_solution}
\end{equation}%
Notice that, in order to compute (\ref{Q1_and_Q2_second_solution}), we took
different signs of $\kappa $ on the two regions $x+y>0$ and $x+y<0$
(however, these signs do not enter in the GPE-which depends on $\kappa ^{2}$%
-and so it is satisfied on both regions).

\begin{figure}[tbp]
\label{fig_vector} \centering
\includegraphics[scale=0.5]{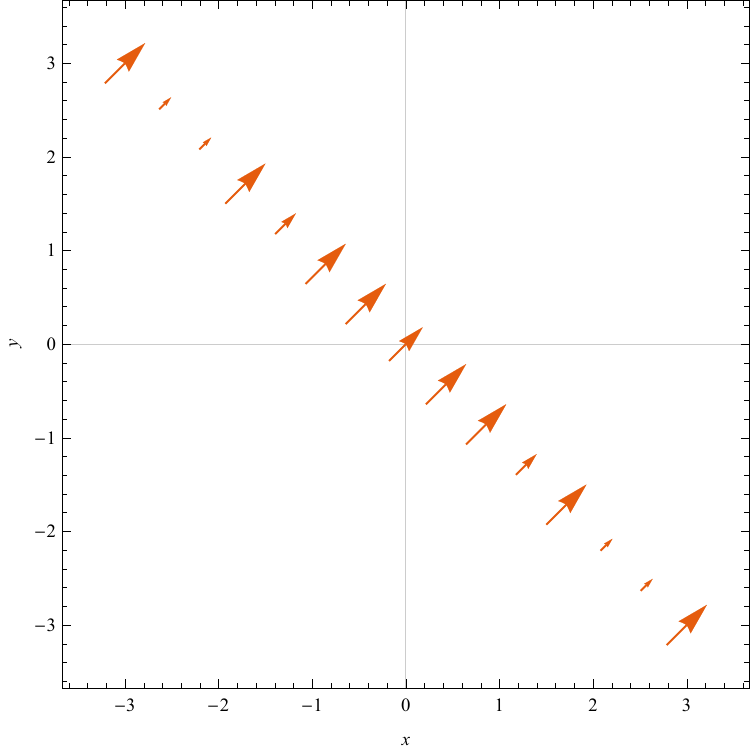}
\caption{ The vector plot of $\vec{J}$, where it is always perpendicular to
the line $\protect\sqrt{2}\,A- \,\protect\kappa \,(x\,\cos \,(3S_{0}-\protect%
\gamma )+y\sin\,(3S_{0}-\protect\gamma ))=0$. In this case, $A=0$, $\protect%
\kappa=1$ and $3S_{0}+\protect\gamma=\frac{\protect\pi}{4}$.}
\label{fig_vector}
\end{figure}

\begin{comment}
On the other hand, $Q_{2}$ is non-trivial:
\begin{equation*}
Q_{2}=-\frac{2\hbar \,\kappa ^{2}\,\pi \,R\,\cos (3S_{0}-\gamma -\theta )}{%
M\,(\sqrt{2}\,A\mp \,\kappa \,(x\,\cos \,(3S_{0}-\gamma )+y\sin
\,(3S_{0}-\gamma )))^{3}}\;.
\end{equation*}%
\textbf{XXX Aquí $Q_{2}$ no se puede calcular con la fórmula anterior porque no está definida en toda una línea, y no podemos integrar en un círculo!XXX} Here is an explicit instance where $Q_{1}$ is zero but the second charge is not, showing they represent different physical quantities.
\end{comment}

\section{Amplitude-phase Relation}

\label{sec_AP_rel}

Besides the construction of the exact solutions of GPE described in the
previous sections, the BPS equations allow us to find a closed formula for
the amplitude $\rho (r,\theta )$ in terms of the phase $S(r,\theta )$ of the
wave function $\Phi (r,\theta )$\textit{\ valid for all solutions of the BPS
equations}. In other words, the BPS equations reduce the unknown functions
of the GPE system from two to just one. Indeed, $\rho $ can be derived in
terms of $S$ and its derivatives from the following relation (see Appendix %
\ref{appendix_amplitude_phase_relation} for details)
\begin{equation}
\Delta\,S = \sqrt{2}\,\rho\,\left(\cos (3 S - \gamma - \theta)\frac{%
\partial\,S}{\partial r}+\frac{\sin (3 S - \gamma - \theta)}r\frac{%
\partial\,S}{\partial \theta}\right) \;.  \label{amplitude_phase_relation}
\end{equation}
One consequence of the above relation is that when $S$ is a harmonic
function (so that $\Delta \,S\approx 0$) then $\rho $ must also be close to
zero: $\rho \approx 0$. This conclusion is physically correct as close to
the position of any vortex $S\sim \arctan \frac{y}{x}$ (where the vortex
under investigation has been taken as the origin of the coordinates system).
Indeed, $\arctan \frac{y}{x}$ is harmonic (excluding the origin itself) so
that, close to the origin, $\rho $ must be vanishingly small as expected.
However, this conclusion is not valid anymore once $3S-\gamma -\theta =0$
and $\frac{\partial \,S}{\partial r}=0$, and this is why our solutions (\ref%
{solGPE1}) and (\ref{solGPE2}) are not physically valid at the origin, as $%
\rho $ does not go to zero when $r\rightarrow 0$. Excluding these cases,
once $\rho $ is expressed in terms of $S$ and its derivatives, one can
derive a single master equation for $S$ whose solution can represent
multi-solitonic configurations. We will revisit the analysis of this master
equation in a future publication.

\begin{comment}
\section{Imaginary Chemical Potential (?)}

It is also interesting the BPS equations can be slightly modified to get GPEs with %an imaginary chemical potential.
\begin{eqnarray}
\partial _{x}\phi _{1}+\partial _{y}\phi _{2} + A &=&\frac{\kappa }{\sqrt{2}}\left(\phi_{1}^{2}-\phi _{2}^{2}\right) \;,
\label{che_GP_1} \\
\partial _{y}\phi _{1}-\partial _{x}\phi _{2} + B &=&\sqrt{2}\,\kappa\,\phi _{1}\phi _{2}\;.
\label{che_GP_2}
\end{eqnarray}

\textbf{Fabrizio, chequea, pero no me da potencial químico ni real ni imgainario, me da algo como}
$$\Delta\,\Phi + (A+\ii\,B)\sqrt{2}\,\kappa\overline{\Phi} - \kappa^{2}|\Phi|^{2}\Phi=0 \;,$$
\textbf{que no es un potencial químico sino que es una constante (que puede ser compleja) por el \underline{conjugado} de $\Phi$.}
\end{comment}

\section{Conclusion and Final Remarks}

\label{sec_Conclution}

\label{sec-6} %%%%%%%%%%%%%%%%%%%%%%%%%%%%%%%%%%%%%%%%%%%%

The present manuscript presents a novel BPS bound for the GPE in two spatial
dimensions: when such a first-order BPS system is satisfied, the
second-order GPE equation is also satisfied. The configurations which
saturate the bound may represent configurations with fractional vorticity.
Quite remarkably, the energy can be computed exactly in terms of a suitable
topological charge. The topological charge is the combination of two
boundary terms. One is related to the vorticity, while the second is a novel
charge that considers the shapes of the BPS configurations. In conclusion,
the main achievement of the present formalism is to replace (in the case of
static configurations) the second-order GPE with a first-order BPS system
that is much simpler to solve numerically and analytically. The numerical
solutions of such a BPS system will be discussed in a future publication.

The main results of this manuscript can be generalized to take into account
terms proportional to $|\Phi|^{2}$ (chemical potential kind of terms) and $%
|\Phi|^{6}$, which takes into account interactions of different solutions
with fractional vorticity. All this is explored in detail in \cite%
{Canfora2025}.

The analytic solutions of the GPE found using the BPS system are the first
analytic examples of configurations with fractional vorticity and with a
non-trivial radial profile defined within an annulus. The BPS system also
allows us to derive a closed expression for the amplitude of the wave
function in terms of the phase, which is valid for any solution of the BPS
system. Such expression is the basis of the analysis of the multi-solitonic
configurations. Moreover, the availability of a first-order BPS system
allows us to derive relevant exact information on the hydrodynamic behavior
of the superfluid. We will come back to this issue in a future publication.

%%%%%%%%%%%%%%%%%%%%%%%%%%%%%%%%%%%%%%%%%%%%%

\section*{Acknowledgements}

%%%%%%%%%%%%%%%%%%%%%%%%%%%%%%%%%%%%%%%%%%%%%

F. C. has been funded by FONDECYT Grants No. 1240048, 1240043 and 1240247.
P.~P. gladly acknowledge support from Charles University Research Center
(UNCE 24/SCI/016).

\begin{appendices}

\section{GP solutions with BPS}

\label{appendix_solutions_without_rotation}

Let us start by considering the BPS equations given in (\ref{BPSGPE1}) and (%
\ref{BPSGPE2})
\begin{eqnarray}
\partial _{x}\phi _{1}+\partial _{y}\phi _{2} &=&\frac{\kappa}{\sqrt{2}}%
\left(\phi_{1}^{2}-\phi _{2}^{2}\right) \;,  \label{fo_GP_1} \\
\partial _{y}\phi _{1}-\partial _{x}\phi _{2} &=&\sqrt{2}\,\kappa\,\phi
_{1}\phi _{2}\;.  \label{fo_GP_2}
\end{eqnarray}

One can directly obtain the GPEs from this BPS system by direct computation.
However, it is remarkable that the GPEs can be obtained for this more
general system
\begin{eqnarray}
\partial _{x}\phi _{1}+\partial _{y}\phi _{2} &=&\frac{a_{1}}{2}%
\left(\phi_{1}^{2}-\phi _{2}^{2}\right) + a_{2}\,\phi_{1}\,\phi_{2} \;,
\label{mod_BPS_1} \\
\partial _{y}\phi _{1}-\partial _{x}\phi _{2} &=&a_{1}\,\phi _{1}\phi _{2} -\frac{a_{2}}{2}%
\left(\phi_{1}^{2}-\phi _{2}^{2}\right) \;.
\label{mod_BPS_2}
\end{eqnarray}
where $a_{1}$ and $a_{2}$ are two real constants that fulfill the condition
\begin{equation}  \label{a1_a2_cond}
a_{1}^{2} + a_{2}^{2}=2\kappa^{2} \;.
\end{equation}
As can be directly checked, the system (\ref{fo_GP_1})--(\ref{fo_GP_2}) is
obtained from (\ref{mod_BPS_1})--(\ref{mod_BPS_2}) when $a_{1}=\sqrt{2}%
\,\kappa$ and $a_{2}=0$. The condition (\ref{a1_a2_cond}) can be written
through a real parameter $\gamma$ through
\begin{equation}  \label{a1_a2_gamma}
a_{1}=\sqrt{2}\,\kappa\,\cos\,\gamma \;, \quad \mbox{and} \quad a_{2}=\sqrt{2%
}\,\kappa\,\sin\,\gamma \;.
\end{equation}

Now, take the fields in the amplitude-phase form (\ref{amplitude-phase}),
i.e.,
\begin{eqnarray}  \label{rho_S}
\phi_{1}(r,\theta)&=&\rho(r,\theta)\cos\,S(r,\theta) \;,  \notag \\
\phi_{2}(r,\theta)&=&\rho(r,\theta)\sin\,S(r,\theta) \;.
\end{eqnarray}
where the polar coordinates are
\begin{eqnarray}  \label{polar_coordinates}
r&=&\sqrt{x^{2}+y^{2}} \;,  \notag \\
\theta&=&\arctan(\frac{y}{x}) \;.
\end{eqnarray}
Therefore, the right-hand side of (\ref{mod_BPS_1}) and (\ref{mod_BPS_2})
can be written as
\begin{eqnarray*}
\frac{a_{1}}{2}\left(\phi_{1}^{2}-\phi _{2}^{2}\right) +
a_{2}\,\phi_{1}\,\phi_{2} &=& \frac{\kappa}{\sqrt{2}}\rho^{2}(\cos\,\gamma\,%
\cos^{2}S-\cos\,\gamma\,\sin^{2}S+2\sin\,\gamma\,\cos\,S\sin\,S) \\
&=& \frac{\kappa}{\sqrt{2}}\rho^{2}\,\cos\,(2S-\gamma) \\
-\frac{a_{2}}{2}\left(\phi_{1}^{2}-\phi _{2}^{2}\right) + a_{1}\,\phi
_{1}\phi _{2} &=& \frac{\kappa}{\sqrt{2}}(-\sin\,\gamma\,\cos^{2}S+\sin\,%
\gamma\,\sin^{2}S+2\cos\,\gamma\,\cos\,S\sin\,S) \\
&=& \frac{\kappa}{\sqrt{2}}\rho^{2}\,\sin\,(2S-\gamma) \\
\end{eqnarray*}

Using the change of variables formulae, we get for the left-hand side of (\ref%
{mod_BPS_1}) and (\ref{mod_BPS_2}),
\begin{eqnarray*}
\frac{\partial\,\Phi_{1}}{\partial x} &=& \frac{\partial\,r}{\partial x}%
\frac{\partial\,\Phi_{1}}{\partial r} + \frac{\partial\,\theta}{\partial x}%
\frac{\partial\,\Phi_{1}}{\partial \theta} = \frac{\partial\,\rho}{\partial r%
}\,\cos\theta\,\cos\,S - \rho\,\frac{\partial\,S}{\partial r}%
\,\cos\theta\,\sin\,S -\frac{\partial\,\rho}{\partial \theta}\,\frac{%
\sin\theta\,\cos\,S}{r} + \rho\frac{\partial\,S}{\partial \theta}\,\frac{%
\sin\theta\,\sin\,S}{r} \\
\frac{\partial\,\Phi_{2}}{\partial y} &=& \frac{\partial\,r}{\partial y}%
\frac{\partial\,\Phi_{2}}{\partial r} + \frac{\partial\,\theta}{\partial y}%
\frac{\partial\,\Phi_{2}}{\partial \theta} = \frac{\partial\,\rho}{\partial r%
}\,\sin\theta\,\sin\,S + \rho\,\frac{\partial\,S}{\partial r}%
\,\sin\theta\,\cos\,S +\frac{\partial\,\rho}{\partial \theta}\,\frac{%
\cos\theta\,\sin\,S}{r} + \rho\frac{\partial\,S}{\partial \theta}\,\frac{%
\cos\theta\,\cos\,S}{r} \\
\frac{\partial\,\Phi_{1}}{\partial y} &=& \frac{\partial\,r}{\partial y}%
\frac{\partial\,\Phi_{1}}{\partial r} + \frac{\partial\,\theta}{\partial y}%
\frac{\partial\,\Phi_{1}}{\partial \theta} = \frac{\partial\,\rho}{\partial r%
}\,\sin\theta\,\cos\,S - \rho\,\frac{\partial\,S}{\partial r}%
\,\sin\theta\,\sin\,S+\frac{\partial\,\rho}{\partial \theta}\,\frac{%
\cos\theta\,\cos\,S}{r} - \rho\frac{\partial\,S}{\partial \theta}\,\frac{%
\cos\theta\,\sin\,S}{r} \\
\frac{\partial\,\Phi_{2}}{\partial x} &=& \frac{\partial\,r}{\partial x}%
\frac{\partial\,\Phi_{2}}{\partial r} + \frac{\partial\,\theta}{\partial x}%
\frac{\partial\,\Phi_{2}}{\partial \theta} = \frac{\partial\,\rho}{\partial r%
}\,\cos\theta\,\sin\,S + \rho\,\frac{\partial\,S}{\partial r}%
\,\cos\theta\,\cos\,S - \frac{\partial\,\rho}{\partial \theta}\,\frac{%
\sin\theta\,\sin\,S}{r} - \rho\frac{\partial\,S}{\partial \theta}\,\frac{%
\sin\theta\,\cos\,S}{r} \;. \\
\end{eqnarray*}
Then, by substituting in (\ref{mod_BPS_1}) and (\ref{mod_BPS_2}),
\begin{eqnarray*}
\frac{\partial\,\rho}{\partial r}\left(\cos\theta\,\cos\,S +
\sin\theta\,\sin\,S\right) &+& \rho\frac{\partial\,S}{\partial r}%
\left(-\cos\theta\,\sin\,S+\sin\theta\,\cos\,S\right) + \frac{1}{r}\frac{%
\partial\,\rho}{\partial \theta}\left(-\sin\theta\,\cos\,S+\cos\theta\,\sin%
\,S\right) + \\
&&\frac{\rho}{r}\frac{\partial\,S}{\partial \theta}\left(\sin\theta\,\sin%
\,S+\cos\theta\,\cos\,S\right) = \frac{\kappa}{\sqrt{2}}\rho^{2}\cos\,(2S-%
\gamma) \;, \\
\frac{\partial\,\rho}{\partial r}\left(\sin\theta\,\cos\,S -
\cos\theta\,\sin\,S\right) &+& \rho\frac{\partial\,S}{\partial r}%
\left(-\sin\theta\,\sin\,S-\cos\theta\,\cos\,S\right) + \frac{1}{r}\frac{%
\partial\,\rho}{\partial \theta}\left(\cos\theta\,\cos\,S+\sin\theta\,\sin%
\,S\right) + \\
&&\frac{\rho}{r}\frac{\partial\,S}{\partial \theta}\left(-\cos\theta\,\sin%
\,S+\sin\theta\,\cos\,S\right) = \frac{\kappa}{\sqrt{2}}\rho^{2}\sin\,(2S-%
\gamma)\;.
\end{eqnarray*}
Therefore, by using the angle sum formulae for sines and cosines,
\begin{eqnarray*}
\frac{\partial\,\rho}{\partial r}\,\cos(\theta-S) + \rho\frac{\partial\,S}{%
\partial r}\sin(\theta-S) - \frac{1}{r}\frac{\partial\,\rho}{\partial \theta}%
\sin(\theta-S) + \frac{\rho}{r}\frac{\partial\,S}{\partial \theta}%
\cos(\theta-S) = \frac{\kappa}{\sqrt{2}}\rho^{2}\cos\,(2S-\gamma) \;, \\
\frac{\partial\,\rho}{\partial r}\sin(\theta-S) - \rho\frac{\partial\,S}{%
\partial r}\cos(\theta-S) + \frac{1}{r}\frac{\partial\,\rho}{\partial\theta}%
\cos(\theta-S) + \frac{\rho}{r}\frac{\partial\,S}{\partial \theta}%
\sin(\theta-S) = \frac{\kappa}{\sqrt{2}}\rho^{2}\sin\,(2S-\gamma) \;,
\end{eqnarray*}
and, by rearranging terms,
\begin{eqnarray*}
\left[\frac{\partial\,\rho}{\partial r}+\frac{\rho}{r}\frac{\partial\,S}{%
\partial \theta}\right]\,\cos(\theta-S) + \left[\rho\frac{\partial\,S}{%
\partial r}- \frac{1}{r}\frac{\partial\,\rho}{\partial \theta}\right]%
\,\sin(\theta-S) = \frac{\kappa}{\sqrt{2}}\rho^{2}\cos\,(2S-\gamma) \;, \\
\left[\frac{\partial\,\rho}{\partial r}+\frac{\rho}{r}\frac{\partial\,S}{%
\partial \theta}\right]\,\sin(\theta-S) - \left[\rho\frac{\partial\,S}{%
\partial r}- \frac{1}{r}\frac{\partial\,\rho}{\partial \theta}\right]%
\,\cos(\theta-S) = \frac{\kappa}{\sqrt{2}}\rho^{2}\sin\,(2S-\gamma)\;.
\end{eqnarray*}
Let us define the functions
\begin{eqnarray*}
\alpha(r,\theta) &\equiv& \frac{\partial\,\rho}{\partial r}+\frac{\rho}{r}%
\frac{\partial\,S}{\partial \theta} \;, \\
\beta(r,\theta) &\equiv& \rho\frac{\partial\,S}{\partial r}- \frac{1}{r}%
\frac{\partial\,\rho}{\partial \theta} \;,
\end{eqnarray*}
then,
\begin{eqnarray}  \label{alpha_beta_rel}
\alpha\,\cos(\theta-S) + \beta\,\sin(\theta-S) = \frac{\kappa}{\sqrt{2}}%
\rho^{2}\cos\,(2S-\gamma) \;,  \notag \\
\alpha\,\sin(\theta-S) - \beta\,\cos(\theta-S) = \frac{\kappa}{\sqrt{2}}%
\rho^{2}\sin\,(2S-\gamma) \;.
\end{eqnarray}
By defining $\delta=\arctan(\frac{\beta}{\alpha})$, along with
\begin{eqnarray*}
\alpha&=&\frac{\kappa}{\sqrt{2}}\rho^{2}\cos\,\delta \;, \\
\beta&=&\frac{\kappa}{\sqrt{2}}\rho^{2}\sin\,\delta \;,
\end{eqnarray*}
and performing some algebraic manipulations, we get from (\ref%
{alpha_beta_rel}),
\begin{eqnarray*}
\alpha^{2} + \beta^{2} = \frac{\kappa^{2}}{2}\rho^{4} \;, \\
\tan(\theta-S-\delta) = \tan\,(2S-\gamma) \;.
\end{eqnarray*}
The last equation is possible if $\delta = \theta - 3S + \gamma $.
Therefore,
\begin{eqnarray*}
\alpha&=&\frac{\kappa}{\sqrt{2}}\rho^{2}\cos\,(\theta-3S+\gamma) \;, \\
\beta&=&\frac{\kappa}{\sqrt{2}}\rho^{2}\sin\,(\theta-3S+\gamma) \;,
\end{eqnarray*}
or,
\begin{eqnarray*}
\frac{\partial\,\rho}{\partial r}+\frac{\rho}{r}\frac{\partial\,S}{\partial
\theta} &=& \frac{\kappa}{\sqrt{2}}\rho^{2}\cos\,(\theta-3S+\gamma) \;, \\
\rho\frac{\partial\,S}{\partial r}- \frac{1}{r}\frac{\partial\,\rho}{%
\partial \theta}&=& \frac{\kappa}{\sqrt{2}}\rho^{2}\sin\,(\theta-3S+\gamma)
\;,
\end{eqnarray*}
This equation system is very general. Let us suppose $\frac{\partial\,S}{\partial r}=0$.
\begin{eqnarray*}
\frac{\partial\,\rho}{\partial r}+\frac{\rho}{r}\frac{\partial\,S}{\partial
\theta} &=& \frac{\kappa}{\sqrt{2}}\rho^{2}\cos\,(\theta-3S+\gamma) \;, \\
- \frac{1}{r}\frac{\partial\,\rho}{\partial \theta}&=& \frac{\kappa}{\sqrt{2}%
}\rho^{2}\sin\,(\theta-3S+\gamma) \;,
\end{eqnarray*}
or,
\begin{eqnarray*}
-\frac{\partial\,u}{\partial r}+\frac{u}{r}\frac{\partial\,S}{\partial \theta%
} &=& \frac{\kappa}{\sqrt{2}}\,\cos\,(\theta-3S+\gamma) \;, \\
\frac{1}{r}\frac{\partial\,u}{\partial \theta}&=& \frac{\kappa}{\sqrt{2}}%
\sin\,(\theta-3S+\gamma) \;,
\end{eqnarray*}
where we defined $u\equiv\frac{1}{\rho}$.

By deriving the first equation with respect to $\theta$ and the second one
with respect to $r$, we get,
\begin{eqnarray*}
-\frac{\partial^{2}\,u}{\partial \theta \, \partial r}+\frac{1}{r}\frac{%
\partial\,u}{\partial\,\theta}\frac{\partial\,S}{\partial \theta} + \frac{u}{%
r}\frac{\partial^{2}\,S}{\partial \theta^{2}} &=& -(1-3\frac{\partial\,S}{%
\partial\,\theta})\frac{\kappa}{\sqrt{2}}\,\sin\,(\theta-3S+\gamma) \;, \\
-\frac{1}{r^{2}}\frac{\partial\,u}{\partial \theta} + \frac{1}{r}\frac{%
\partial^{2}\,u}{\partial r \, \partial \theta}&=& 0\;.
\end{eqnarray*}
By substituting $\frac{\partial^{2}\,u}{\partial r \, \partial \theta}$ from
the second equation into the first one,
\begin{equation*}
\frac{1}{r}\frac{\partial\,u}{\partial \theta}\left[-1+\frac{\partial\,S}{%
\partial \theta}\right] + \frac{u}{r}\frac{\partial^{2}\,S}{\partial
\theta^{2}} = -(1-3\frac{\partial\,S}{\partial\,\theta})\frac{\kappa}{\sqrt{2%
}}\,\sin\,(\theta-3S+\gamma)\;,
\end{equation*}
or,
\begin{eqnarray*}
&& \frac{\kappa}{\sqrt{2}}\,\sin\,(\theta-3S+\gamma)\left[-1+\frac{%
\partial\,S}{\partial \theta}\right] + \frac{u}{r}\frac{\partial^{2}\,S}{%
\partial \theta^{2}} = -\frac{\kappa}{\sqrt{2}}\,(1-3\frac{\partial\,S}{%
\partial\,\theta})\,\sin\,(\theta-3S+\gamma) \\
&&\quad \Rightarrow \quad \frac{u}{r}\frac{\partial^{2}\,S}{\partial
\theta^{2}} = \sqrt{2}\,\kappa\,\frac{\partial\,S}{\partial\,\theta}%
\,\sin\,(\theta-3S+\gamma)\;.
\end{eqnarray*}
There are at least two direct ways to fulfil this equation, but there could be more.

\subsection{First kind of solution: $\protect\delta=\protect\theta-3S+%
\protect\gamma=0$}

If $\theta-3S+\gamma=0$, then $S(\theta)=\frac{\theta}{3} + \frac{\gamma}{3}
$. So,
\begin{eqnarray*}
-\frac{\partial\,u}{\partial r}+\frac{u}{3r} &=& \frac{\kappa}{\sqrt{2}} \;,
\\
\frac{1}{r}\frac{\partial\,u}{\partial \theta}&=& 0 \;,
\end{eqnarray*}
implying $u$ does not depend on $\theta$. Then $u$ should satisfy the ODE
\begin{equation}
-u^{\prime }+\frac{u}{3r} = \frac{\kappa}{\sqrt{2}} \;.
\end{equation}
This is satisfied if $u(r)=u_{H}(r)+u_{P}(r)$, where $u_{H}$ is a solution of
the homogeneous equation
\begin{equation}
-u^{\prime }+\frac{u}{3r} =0\;,
\end{equation}
whose general expression is $u_{H}(r)=A\,r^{\frac{1}{3}}$, being $A$ an
integration constant. A particular solution could be written as $%
u_{P}(r)=B\,r^{q}$, where
\begin{equation}
-B\,q\,r^{q-1}+\frac{B}{3}r^{q-1} = \,\frac{\kappa}{\sqrt{2}} \quad
\Rightarrow \quad q=1 \quad \mbox{and} \quad B=-\frac{3}{2\sqrt{2}}\kappa\;.
\end{equation}
Then, $u_{P}(r)=-\,\frac{3}{2\sqrt{2}}\kappa\,r$, and $u(r)=A\,r^{\frac{1}{3}%
}-\frac{3}{2\sqrt{2}}\kappa\,r$, giving us,
\begin{equation}\label{rho_first_sol}
\rho(r)=\frac{2\sqrt{2}}{2\sqrt{2}\,A\,r^{\frac{1}{3}}-\,3\kappa\,r} \;.
\end{equation}
Finally, we get the two fields
\begin{eqnarray*}
\phi_{1}(r,\theta) &=& \frac{2\sqrt{2}\,\cos\,\left(\frac{\theta}{3}+\frac{%
\gamma}{3}\right)}{2\sqrt{2}\,A\,r^{\frac{1}{3}}-\,3\kappa\,r} \;, \\
\phi_{2}(r,\theta) &=& \frac{2\sqrt{2}\,\sin\,\left(\frac{\theta}{3}+\frac{%
\gamma}{3}\right)}{2\sqrt{2}\,A\,r^{\frac{1}{3}}-\,3\kappa\,r} \;.
\end{eqnarray*}
By redefining the parameter $\theta_{0}=\frac{\gamma}{3}$ we obtain the
solution
\begin{eqnarray*}
\phi_{1}(r,\theta) &=& \frac{2\sqrt{2}\,\cos\,\left(\frac{\theta}{3}%
+\theta_{0}\right)}{2\sqrt{2}\,A\,r^{\frac{1}{3}}-\,3\kappa\,r} \;, \\
\phi_{2}(r,\theta) &=& \frac{2\sqrt{2}\,\sin\,\left(\frac{\theta}{3}%
+\theta_{0}\right)}{2\sqrt{2}\,A\,r^{\frac{1}{3}}-\,3\kappa\,r}\;.
\end{eqnarray*}

Take into account that this solution could diverge, apart from the origin, when $%
\sqrt{2}\,A\,r^{\frac{1}{3}}-\,3\kappa\,r=0$.

\subsection{Second kind of solution: $\frac{\partial\,S}{\partial\,\protect%
\theta}=0$}

If $\frac{\partial\,S}{\partial\,\theta}=0$, then $S(\theta)=S_{0}$. Therefore,
\begin{eqnarray*}
-\frac{\partial\,u}{\partial r}&=& \,\frac{\kappa}{\sqrt{2}}\,\cos(\theta -
3\,S_{0}+ \gamma)\;, \\
\frac{1}{r}\frac{\partial\,u}{\partial \theta}&=& \,\frac{\kappa}{\sqrt{2}}%
\,\sin(\theta - 3\,S_{0} + \gamma) \;,
\end{eqnarray*}
The second equation suggests that
\begin{equation*}
u(r,\theta)=-\,\frac{\kappa}{\sqrt{2}} \,r\,\cos(\theta - 3\,S_{0} + \gamma)
+ u_{0}(r) \;,
\end{equation*}
where $u_{0}(r)$ only depends on $r$. By substituting this into the first
equation, we get
\begin{equation*}
-u_{0}^{\prime }=0 \quad \Rightarrow \quad u_{0}=A\;, \quad
\mbox{where $A$
is an integration constant}\;.
\end{equation*}
Therefore,
\begin{equation*}
u(r,\theta)=-\,\frac{\kappa}{\sqrt{2}} \,r\,\cos(\theta - 3\,S_{0} + \gamma)
+ A \;,
\end{equation*}
and,
\begin{equation*}
\rho(r,\theta)=\frac{\sqrt{2}}{\sqrt{2}\,A-\,\kappa\,r\,\cos(\theta -
3\,S_{0} + \gamma)} \;.
\end{equation*}
Finally, we get the two fields
\begin{eqnarray*}
\phi_{1}(r,\theta) &=& \frac{\sqrt{2}\,\cos\,S_{0}}{\sqrt{2}%
\,A-\,\kappa\,r\,\cos(\theta - 3\,S_{0}+\gamma)} \;, \\
\phi_{2}(r,\theta) &=& \frac{\sqrt{2}\,\sin\,S_{0}}{\sqrt{2}%
\,A-\,\kappa\,r\,\cos(\theta - 3\,S_{0}+\gamma)} \;.
\end{eqnarray*}
%Take into account this solution diverges when $\cos(\theta - 3\,S_{0}+\gamma)= \,\frac{A}{\kappa\,r}$, so if $r\ll \frac{A}{\kappa}$, it is a well-behaved soultion.
By coming back to Cartesian coordinates,
\begin{eqnarray*}
\phi_{1}(x,y) &=& \frac{\sqrt{2}\,\cos\,S_{0}}{\sqrt{2}\,A-\,\kappa\,(x\,%
\cos\,(3S_{0}-\gamma)+y\sin\,(3S_{0}-\gamma))}\;, \\
\phi_{2}(x,y) &=& \frac{\sqrt{2}\,\sin\,S_{0}}{\sqrt{2}\,A-\,\kappa\,(x\,%
\cos\,(3S_{0}-\gamma)+y\sin\,(3S_{0}-\gamma))} \;.
\end{eqnarray*}

\section{Deduction of an amplitude-phase relation for BPS solutions}

\label{appendix_amplitude_phase_relation}

Here, we will show that a solution of the BPS equation does not have a totally independent amplitude $\rho$ and phase $S$.

Suppose $\phi_{1}$ and $\phi_{2}$ are two solutions of BPS equations (\ref{mod_BPS_1})--(\ref{mod_BPS_2}), and we know the phase function $S(r,\theta)$.
By using the same trigonometric formulae below equation (\ref%
{polar_coordinates}), we get then,
\begin{eqnarray*}
(\frac{\partial\,\rho}{\partial x} + \rho\,\frac{\partial\,S}{\partial y}%
)\,\cos\,S + (\frac{\partial\,\rho}{\partial y} - \rho\,\frac{\partial\,S}{%
\partial x})\,\sin\,S &=& \frac{\kappa}{\sqrt{2}}\,\rho^{2}\,\cos\,(2S-%
\gamma) \;, \\
(\frac{\partial\,\rho}{\partial y} - \rho\,\frac{\partial\,S}{\partial x}%
)\,\cos\,S - (\frac{\partial\,\rho}{\partial x} + \rho\,\frac{\partial\,S}{%
\partial y})\,\sin\,S &=& \frac{\kappa}{\sqrt{2}}\,\rho^{2}\,\sin\,(2S-%
\gamma) \;. \\
\end{eqnarray*}
By changing the variable $u=\frac{1}{\rho}$, these equations are
\begin{eqnarray*}
(-\frac{\partial\,u}{\partial x} + u\,\frac{\partial\,S}{\partial y}%
)\,\cos\,S + (-\frac{\partial\,u}{\partial y} - u\,\frac{\partial\,S}{%
\partial x})\,\sin\,S &=& \frac{\kappa}{\sqrt{2}}\,\cos\,(2S- \gamma) \;, \\
(-\frac{\partial\,u}{\partial y} - u\,\frac{\partial\,S}{\partial x}%
)\,\cos\,S - (-\frac{\partial\,u}{\partial x} + u\,\frac{\partial\,S}{%
\partial y})\,\sin\,S &=& \frac{\kappa}{\sqrt{2}}\,\sin\,(2S- \gamma) \;. \\
\end{eqnarray*}
As we know the function $S(x,y)$, these two equations must be satisfied by
the only unknown function $u(x,y)$, implying some constraints. Let us write
\begin{eqnarray}  \label{general_system}
A_{1}\,\frac{\partial\,u}{\partial x} + B_{1}\,\frac{\partial\,u}{\partial y}
+ C_{1}\,u&=& F_{1} \;, \\
A_{2}\,\frac{\partial\,u}{\partial x} + B_{2}\,\frac{\partial\,u}{\partial y}
+ C_{2}\,u&=& F_{2} \;, \\
\end{eqnarray}
where
\begin{eqnarray*}
A_{1} = -\cos\,S\;, \quad B_{1}=-\sin\,S\;, \quad C_{1}=\frac{\partial\,S}{%
\partial y}\cos\,S - \frac{\partial\,S}{\partial x}\sin\,S\;, & & F_{1} = \frac{\kappa}{\sqrt{2}}\,\cos\,(2S- \gamma)\;, \\
A_{2}=\sin\,S\;,\ \quad B_{2}=-\cos\,S\, \quad C_{2}= -\frac{\partial\,S}{%
\partial x}\cos\,S - \frac{\partial\,S}{\partial y}\sin\,S\;, & & F_{2} =
\frac{\kappa}{\sqrt{2}}\,\sin\,(2S- \gamma) \;, \\
\end{eqnarray*}
By dividing the first equation by $A_{1}$ and the second by $A_{2}$, we get,
\begin{eqnarray*}
\frac{\partial\,u}{\partial x} + \tilde{B}_{1}\,\frac{\partial\,u}{\partial y%
} + \tilde{C}_{1}\,u&=& \tilde{F}_{1} \;, \\
\frac{\partial\,u}{\partial x} + \tilde{B}_{2}\,\frac{\partial\,u}{\partial y%
} + \tilde{C}_{2}\,u&=& \tilde{F}_{2} \;, \\
\end{eqnarray*}
and, by substracting the first to the second,
\begin{equation*}
(\tilde{B}_{2}-\tilde{B}_{1})\,\frac{\partial\,u}{\partial y} + (\tilde{C}%
_{2}-\tilde{C}_{1})\,u = \tilde{F}_{2} - \tilde{F}_{1} \;.
\end{equation*}
It can be shown
\begin{eqnarray*}
\tilde{B}_{2}-\tilde{B}_{1} &=& -\cot\,S - \tan\,S\;, \\
\tilde{C}_{2}-\tilde{C}_{1} &=& -\frac{\partial\,S}{\partial x}%
\,(\tan\,S+\cot\,S)\;, \\
\tilde{F}_{2}-\tilde{F}_{1} &=&\frac{\kappa}{\sqrt{2}}\,\kappa\left(\frac{\cos\,(2S- \gamma)%
}{\cos\,S} + \frac{\sin\,(2S- \gamma)}{\sin\,S}\right)\;.
\end{eqnarray*}
Therefore,
\begin{equation}  \label{condition_u_y_multivortex}
\frac{\partial\,u}{\partial y} - A(x,y)\,u = B(x,y) \;.
\end{equation}
where we defined the functions
\begin{eqnarray*}
A(x,y)&\equiv&-\frac{\partial\,S}{\partial x}\;, \\
B(x,y)&\equiv& -\frac{\kappa}{\sqrt{2}}\,\sin\,(3S- \gamma)\;.
\end{eqnarray*}
The formal solution for equation (\ref{condition_u_y_multivortex}) is
\begin{equation}  \label{genral_solution_u_y}
u(x,y)=e^{\int_{y_{0}}^{y}\,A(x,y^{\prime })\,dy^{\prime }}\left[ \alpha(x)
+ \int_{y_{0}}^{y}\,e^{-\int_{y_{0}}^{y^{\prime }}\,A(x,y^{\prime \prime
})\,dy^{\prime \prime }}\,B(x,y^{\prime })\,dy^{\prime }\right] \;,
\end{equation}
where $\alpha$ is a function only on $x$.

Now, let us come back to equation (\ref{general_system}), but this time, we
divide the first equation by $B_{1}$ and the second by $B_{2}$,
\begin{eqnarray*}
\tilde{A}_{1}\,\frac{\partial\,u}{\partial x} + \frac{\partial\,u}{\partial y%
} + \tilde{\tilde{C}}_{1}\,u&=& \tilde{\tilde{F}}_{1} \;, \\
\tilde{A}_{2}\,\frac{\partial\,u}{\partial x} +\frac{\partial\,u}{\partial y}
+ \tilde{\tilde{C}}_{2}\,u&=& \tilde{\tilde{F}}_{2} \;, \\
\end{eqnarray*}
and, by subtracting the first to the second,
\begin{equation*}
(\tilde{A}_{2}-\tilde{A}_{1})\,\frac{\partial\,u}{\partial x} + (\tilde{%
\tilde{C}}_{2}-\tilde{\tilde{C}}_{1})\,u = \tilde{\tilde{F}}_{2} - \tilde{%
\tilde{F}}_{1} \;.
\end{equation*}

It can be shown
\begin{eqnarray*}
\tilde{A}_{2}-\tilde{A}_{1} &=& -\cot\,S - \tan\,S\;, \\
\tilde{\tilde{C}}_{2}-\tilde{\tilde{C}}_{1} &=& \frac{\partial\,S}{\partial y%
} \left(\cot\,S + \tan\,S\right)\;, \\
\tilde{\tilde{F}}_{2}-\tilde{\tilde{F}}_{1} &=& \frac{\kappa}{\sqrt{2}}\left(\frac{%
\cos\,(2S-\gamma)}{\sin\,S} - \frac{\sin\,(2S-\gamma)}{\cos\,S}\right)\;.
\end{eqnarray*}
Therefore,
\begin{equation}  \label{condition_u_x_multivortex}
\frac{\partial\,u}{\partial x} - C(x,y)\,u = D(x,y) \;.
\end{equation}
where we defined the functions
\begin{eqnarray*}
C(x,y)&\equiv&\frac{\partial\,S}{\partial y}\;, \\
D(x,y)&\equiv&-\frac{\kappa}{\sqrt{2}}\,\cos\,(3S-\gamma)\;.
\end{eqnarray*}
The formal solution for equation (\ref{condition_u_x_multivortex}) is
\begin{equation}  \label{genral_solution_u_x}
u(x,y)=e^{\int_{x_{0}}^{x}\,C(x^{\prime },y)\,dx^{\prime }} \left[\beta(y) +
\int_{x_{0}}^{x}\,e^{-\int_{x_{0}}^{x^{\prime }}\,C(x^{\prime \prime
},y)\,dx^{\prime \prime }}\,D(x^{\prime },y)\,dx^{\prime }\right] \;,
\end{equation}
where $\beta$ is a function only on $y$.

Thus, functions (\ref{genral_solution_u_y}) and (\ref{genral_solution_u_x})
should be the same.

We can derivate the equation (\ref{condition_u_y_multivortex}) with respect
to $x$ and the equation (\ref{condition_u_x_multivortex}) with respect to $y$
and subtracting,
\begin{eqnarray*}
0 = \frac{\partial u}{\partial y \, \partial x} - \frac{\partial u}{\partial
x \, \partial y} &=& \frac{\partial u}{\partial x}A + u\,\frac{\partial A}{%
\partial x} + \frac{\partial B}{\partial x} - \frac{\partial u}{\partial y}C
- u\,\frac{\partial C}{\partial y} - \frac{\partial D}{\partial y} \\
&=& u\left(\frac{\partial A}{\partial x} - \frac{\partial C}{\partial y}%
\right) + AD-BC + \frac{\partial B}{\partial x} - \frac{\partial D}{\partial
y} \;.
\end{eqnarray*}
where in the last equality, we again used equations (\ref{condition_u_y_multivortex}) and (\ref{condition_u_x_multivortex}). That means,
\begin{equation*}
0= -u\left(\frac{\partial^{2} S}{\partial x^{2}} + \frac{\partial^{2} S}{%
\partial y^{2}}\right)- \frac{\sqrt{2}\,\kappa}{3}\,\left(\frac{%
\partial\,\cos (3 S - \gamma)}{\partial y}-\frac{\partial\,\sin (3 S -
\gamma)}{\partial x}\right) \;,
\end{equation*}
or, by going to the polar coordinates,
\begin{equation}  \label{condition_S}
\Delta\,S = \sqrt{2}\,\rho\,\left(\cos (3 S - \gamma -
\theta)\frac{\partial\,S}{\partial r}+\frac{\sin (3 S - \gamma - \theta)}r%
\frac{\partial\,S}{\partial \theta}\right) \;.
\end{equation}
It is reassuring that the two solutions found in Appendix \ref{appendix_solutions_without_rotation} fulfil the condition (\ref{condition_S}).

%\rc{$\rho'' + \frac{(3A+1)}{r}\,\rho' + \frac{-|\frac{\partial S}{\partial \theta}|^{2} + A\frac{\partial S}{\partial \theta} + 2 A^{2}}{r^{2}}\rho - \frac{\sqrt{2}\,\kappa}{2}\rho^{3}$}

\end{appendices}

%\textbf{\ XX sacar referencias no citatas XX}

%%%%%%%%%%%%%%%%%%%%%%%%%%%%%%%%%%%%%%%%%%%%%%% Bibliography %%%%%%%%%%%%%%%%%%%%%%%%%%%%%%%%%%%%%%%%%%%%%%%%%%%%%%%%%%%%%
\bibliographystyle{utphys}
\bibliography{GPE_paper_biblio}
%%%%%%%%%%%%%%%%%%%%%%%%%%%%%%%%%%%%%%%%%%%%%%%%%%%%%%%%%%%%%%%%%%%%%%%%%%%%%%%%%%%%%%%%%%%%%%%%%%%%%%%%%%%%%%%%%%%%%%%%%%

\end{document}